\documentclass[final,3p,11pt,times]{elsarticle}

\usepackage{amssymb,amsmath,mathtools}
\usepackage{epsfig}

 \biboptions{compress}

\journal{Annals of Physics}

\makeatletter
\def\ps@pprintTitle{%
  \let\@oddhead\@empty
  \let\@evenhead\@empty
  \def\@oddfoot{\reset@font\hfil\thepage\hfil}
  \let\@evenfoot\@oddfoot
}
\makeatother

\usepackage{fancyhdr}
\begin{document}

\begin{frontmatter}



\title{\LARGE
Multichannel framework 
\\
for singular quantum mechanics}


\author[USF]{Horacio E. Camblong\corref{cor}}
\ead{camblong@usfca.edu}

\author[UNLP]{Luis N. Epele}
\ead{epele@fisica.unlp.edu.ar}

\author[UNLP]{Huner Fanchiotti}
\ead{huner@fisica.unlp.edu.ar}

\author[UNLP]{Carlos A. Garc\'{\i}a Canal}
\ead{garcia@fisica.unlp.edu.ar}
 
\author[UH]{Carlos R. Ord\'{o}\~{n}ez}
\ead{ordonez@uh.edu}

 \cortext[cor]{
 Corresponding author.}

  \address[USF]{Department of Physics \& Astronomy,  University of San Francisco,  \\
  San Francisco, CA 94117-1080, United States}
  \address[UNLP]{  Laboratorio de F\'{\i}sica Te\'{o}rica,  Departamento de F\'{\i}sica, IFLP, CONICET, 
  Facultad de Ciencias Exactas, Universidad Nacional de La Plata,  C.C. 67 -- 1900 La Plata, Argentina}
 \address[UH]{Department of Physics, University of Houston, Houston, TX 77204-5506, United States}

\begin{abstract}
A {\em multichannel S-matrix framework\/}
for singular quantum mechanics (SQM)
subsumes the renormalization and self-adjoint extension methods
and resolves
its boundary-condition ambiguities.
In addition to the standard channel
accessible to a distant (``asymptotic'')
observer, 
one supplementary channel opens up 
at each coordinate singularity, where
local outgoing and ingoing 
{\em singularity waves\/} coexist. 
The channels are linked by a fully unitary S-matrix,
which governs all possible scenarios,
including cases with an {\em apparent\/}
nonunitary behavior as viewed from asymptotic distances.

\noindent \hrulefill
\end{abstract}



\end{frontmatter}

\section{Introduction}
\label{sec:intro}

The central construct 
 in quantum scattering processes is the S-matrix.
However,
while the orthodox formulation of the {\em S-matrix\/} is applicable 
 to regular quantum mechanics and quantum field theory,
 its generalization for
{\em singular quantum mechanics\/} 
(SQM)
 remains an open problem.
 In effect, the S-matrix orthodoxy is called into question 
 due to the breakdown of the regular boundary condition at the 
 singularity; this can be described as the 
 loss of discriminating value when two linearly independent solutions remain equally acceptable 
 for attractive singularities~\cite{spector_RMP,Newton,landau:77,perelomov:70,esposito}. 
 In addition, there is the issue of
 the possible emergence of nonunitary solutions~\cite{vogt_wannier,Alliluev_ISP}
  and the singular nature of the bound-state spectrum~\cite{gup:93,mee:64_nar:74}.
 These extreme departures from regular quantum mechanics are evident for strongly singular attractive potentials---see 
 definitions in Refs.~\cite{spector_RMP,Newton},
 and in our Section~\ref{sec:BC_Wannier-waves}.

In this paper we circumvent
the divergence and unitarity challenges posed
by SQM
through a novel {\em multichannel framework\/}
that comprehensively subsumes the
well-established {\em renormalization schemes\/}
of Refs.~\cite{gup:93,HEC:ISP_letter,HEC:DTI,HEC:DTII,HEC:dipole_QM-anomaly,HEC:CQManomaly-CQM,beane:00}
and the method of {\em self-adjoint extensions\/}~\cite{cas:50,mee:64_nar:74}.
The latter is an efficient technique to handle this breakdown and the ensuing indeterminacy 
of the solutions by properly defining the domain of the Hamiltonian to 
guarantee self-adjointness. Our framework does provide a practical implementation of this technique, 
but also extends its usefulness beyond self-adjointness. 
In effect, the generalized framework
consistently includes physical realizations with 
effective {\em absorption\/}~\cite{vogt_wannier,Alliluev_ISP}
or {\em emission\/}
by the singular potential---including, for example,
the absorption of particles by charged nanowires 
(with conformal quantum mechanics)~\cite{den:98,aud:99},
the scattering by polarizable molecules
(with an inverse quartic potential)~\cite{vogt_wannier},
and miscellaneous applications to black holes~\cite{BH_thermo_CQM,semiclassical_BH,Hawking_CQM}
and D-branes~\cite{generic_D-brane_IQP_1,generic_D-brane_IQP_2,generic_D-brane_IQP_3,park_D-brane_IQP}.  

In the proposed framework,
a singularity point is treated as
the opening of a new channel  
in lieu of a boundary condition.
A consequence of this constructive approach
is the existence of associated 
local outgoing and ingoing waves
 near the singularity---the
natural generalization of the solutions   
used for the inverse quartic potential 
in Ref.~\cite{vogt_wannier}
and the inverse square potential 
in Ref.~\cite{Alliluev_ISP}.
In our framework,
the crucial guiding criterion is the 
acceptance, on an equal footing,
of these 
{\em ``singularity waves''\/}
as basic building blocks.
In its final form, two related 
quantities are considered:
\begin{enumerate}
\item
A unitary {\em multichannel S-matrix\/} 
$\left[ {\bf S} \right] $,
which effects a separation of the singular behavior
at a point (for example, $r \sim 0$)
from the longer-range properties of the interaction.

\item
An {\em effective or asymptotic S-matrix\/}
$S_{\rm asymp}$,
which directly yields the
scattering observables
viewed by an asymptotic observer
according to the standard procedures of regular quantum mechanics.
\end{enumerate}
While 
$S_{\rm asymp}$
may fail to be unitary
due to the boundary condition ambiguities,
the S-matrix $\left[ {\bf S} \right] $
is guaranteed to satisfy unitarity for the 
quantum system that 
includes a channel connected to the singularity.
In other words, when the singularity point is 
redefined as external to the given system 
(i.e., removed from the observable physics),  
{\em unitarity\/} is automatically restored
 within the enlarged multichannel system.
Specifically, the  S-matrix $S_{\rm asymp}$
is obtained 
via 
a M\"{o}bius transformation~\cite{Mobius} 
of a complex-valued singularity parameter $\Omega$
(which specifies an auxiliary ``boundary condition''), with coefficients
provided by the multichannel S-matrix
$\left[ {\bf S} \right] $.

We have organized our paper as follows.
In Section~\ref{sec:BC_Wannier-waves}
we define singular potentials and SQM, and thereby
 discuss the boundary condition at the singularity
and derive the existence of local ingoing and outgoing waves.
The multichannel framework and properties of the S-matrix are introduced in
Section~\ref{sec:singularQM_multichannel-Smatrix},
leading to the definition of the {\em effective 
asymptotic S-matrix\/} $S_{\rm asymp}$
in Section~\ref{sec:asymptotic_Smatrix}.
In Section~\ref{sec:particular_cases}
we derive the multichannel S-matrix 
$\left[ {\bf S} \right] $
for conformal quantum mechanics 
(Section~\ref{subsec:CQM})
and for the inverse quartic potential (Section~\ref{subsec:IQP});
for the conformal case, we extensively consider additional features arising 
from its SO(2,1) symmetry. 
The paper concludes 
in Section~\ref{sec:conclusions}
with a comparative discussion of the multichannel S-matrix versus 
earlier approaches to SQM and related physical realizations.
In the appendices, we 
discuss transformation properties and 
 we include
two models that modify conformal
quantum mechanics in the infrared,
verifying the robustness of the framework.
Additional analytic properties and technical subtleties 
of this versatile framework are left for a forthcoming follow-up paper.

\section{Singular interactions:
definition, boundary conditions, and ``singularity waves''}
\label{sec:BC_Wannier-waves}

\subsection{Singular quantum mechanics (SQM): definition}
\label{sec:SQM-def}

To properly set up our novel multichannel framework
in its simplest form,
 we will define singular quantum mechanics by
  focusing our analysis on the cases where 
 a {\em strong definition\/} of the concept of singular potential is in place. This assumes 
 a sufficiently  attractive interaction in the neighborhood of the coordinate singularity. 

In the main body of the literature, singular potentials are defined
within a broader class (weak or generic definition),
 including also the repulsive 
ones~\cite{spector_RMP,Newton}. In all cases, the singular class involves interactions at least as dominant 
as  $1/r^{2}$ (order of the angular-momentum potential) around the singular point, 
which we will call {\em coordinate singularity\/},
and is herein identified as $r=0$ (see exceptions below, after Eq.~(\ref{eq:power-law_potential})).
In other words,  a {\em regular potential\/} is defined as one for which 
$\lim_{r \rightarrow 0} r^2 V ( r ) = 0$; 
and a singular one does not satisfy this condition. 
A {\em (properly) singular potential\/}
is defined by the condition $\lim_{r \rightarrow 0} r^2 V ( r ) = \pm \infty$, where 
the $\pm$ sign corresponds respectively to repulsive/attractive potentials (near the singularity). 
For a {\em marginally singular or transitionally singular potential\/} 
(known as the inverse square potential), the limit above is a finite constant:
$\lim_{r \rightarrow 0} r^2 V ( r ) = \lambda_{\rm ISP}$ 
(this is $p=2$ in Eqs.~(\ref{eq:singular-QM_radial})--(\ref{eq:power-law_potential}) below);
but in the presence of logarithmic singularities, the marginal class is defined by the limit 
$\lim_{r \rightarrow 0} r^{2+\epsilon} V ( r ) = 0, \infty$ for all
$\epsilon \stackrel{ >}{_{ <} } 0$~\cite{spector_RMP}.

For the sake of simplicity,
we will restrict our analysis to 
 any problem 
where the relevant physics is
described by the class of  Schr\"{o}dinger-like equations
\begin{equation}
\left[
\frac{d^{2}}{dr^{2}} 
\,
+ k^{2} 
-
V({\bf r})
-
\frac{ \left( l + \nu \right)^{2} - 1/4
}{r^{2}}
\right]  
u (r) = 0
\;  ,
\label{eq:singular-QM_radial}
\end{equation}
in which
the term
\begin{equation}
V({\bf r}) 
\stackrel{( r \rightarrow 0)}{\sim}
-\frac{ \lambda
}{r^{ p } }
\label{eq:power-law_potential}
 \; ,
 \end{equation}
is the 
{\em 
singular interaction\/}.
The  parameters $l$ and $\nu$  (usually associated with angular momentum and spatial dimensionality)
 are to be adjusted
within a given application (see Section~\ref{sec:SQM-applications}); moreover,
the coordinate singularity may be at a point other than the origin, for example the Schwarzschild radius of 
a spherically symmetric
black hole, causing a mismatch between the power $-2$ of the angular momentum and 
the order of the singular
 potential~\cite{HEC:CQManomaly-CQM,BH_thermo_CQM,semiclassical_BH}.
In most cases, though, the angular momentum is combined at the same order with the inverse square potential.
Thus, our formalism broadly
applies to
any problem that reduces to the limiting form of Eqs.~(\ref{eq:singular-QM_radial})--(\ref{eq:power-law_potential})
as $r \rightarrow 0$.
The potentials of Eq.~(\ref{eq:power-law_potential}) 
are the most commonly occurring in physical applications, including the ones listed in Section~\ref{sec:intro} 
and in Section~\ref{sec:SQM-applications}.
But other functional forms are possible, including those with logarithmic and exponential singularities
at the origin~\cite{spector_RMP}.

In essence, the classification above is motivated by the change in the 
analytic properties of the solutions and the S-matrix for Eq.~(\ref{eq:singular-QM_radial}) 
 when the exponent $p$ crosses the $p=2$ threshold. 
In the language of differential equations~\cite{Ince}, the point $r=0$ is an irregular singular point 
for $p>2$, a regular singular point for $p=2$, and an ordinary (nonsingular) point for $p<2$.
Now, even though the scattering quantities (involving integral equations and related analytical tools) 
have peculiar properties for all generalized singular potentials, 
the observable physics becomes distinctly more dramatic 
(e.g., the ``fall to the center'' phenomenon~\cite{landau:77})
 only in the attractive case, which we will often refer to as the {\em strong coupling\/} regime.
 In effect (see Section~\ref{sec:singularity-waves}),
  all strongly attractive singular potentials yield two linearly independent wave-like 
 solutions that are equally acceptable from a first-principle physical viewpoint~\cite{spector_RMP,Newton}---thus,
  they pose an indeterminacy or ``loss of boundary condition"~\cite{HEC:DTII,Schwartz}.
In this paper, we will not address 
the {\em weak coupling\/} regime, i.e., the repulsive singular interactions; notice that
we still refer to strongly repulsive interactions as belonging to the weak-coupling regime, by abuse of language. 
Of course, assuming that the strong-coupling 
results are  established, these  can then be extended to weak coupling by
an appropriate analytic continuation.
Within this context, the inverse square potential---defining conformal quantum mechanics (CQM)---is
the marginal case, for which there is
 a nonzero critical coupling~\cite{landau:77,perelomov:70,HEC:ISP_letter,HEC:DTII}. 
The subtle issues involved in the extension from the strong to the weak coupling, 
and the existence of a medium-coupling window for CQM~\cite{HEC-SA}
will be discussed 
in a forthcoming follow-up paper, along with additional analytic properties of the S-matrix. 

Another qualification of our proposed framework involves contact (point or zero-range) interactions, 
which include Dirac delta functions and  derivatives. 
These generalized functions are also related to boundary conditions, 
though they represent appropriate limits of finite potentials in actual phenomenological applications. 
In principle, they could be added to our description as subsidiary boundary conditions, 
but they play a role distinctly different from power-law singular interactions. 
In contradistinction, 
 the latter can be appropriately 
called {\em non-contact singular potentials\/},
and they are the main focus of our approach. 
Contact interactions may still be hiding in power-law SQM via the boundary conditions 
and/or as renormalization counterterms, but their presence will not be 
 acknowledged in the absence of additional physical justification.
This distinction, based on physical considerations, 
is of prime importance to properly interpret some of the results associated 
with domains of operators and self-adjoint extensions---an issue 
that is crucial to a proper interpretation of the medium-weak (intermediate) coupling in CQM, 
as discussed in Ref.~\cite{HEC-SA}. 
We will extend our study of analytic properties of our multichannel framework 
in our follow-up paper, where these subtle issues and distinctions will be further dissected.

\subsection{SQM: range of applicability}
\label{sec:SQM-applications}

As mentioned in Section~\ref{sec:intro}, 
SQM can have a wide range of applications.
Thus, depending on the context, 
Eq.~(\ref{eq:singular-QM_radial}) may have very different 
physical interpretations.

For the particular case of nonrelativistic quantum mechanics,
Eq.~(\ref{eq:singular-QM_radial}) is the familiar radial
counterpart of the
ordinary Schr\"{o}dinger equation, generically
derived via 
the representation
$
\Psi ({\bf r})
=  
Y_{l m}( {\bf \Omega})
u(r)/r^{(d-1)/2} $
of the multidimensional wave function 
in $d$  spatial dimensions,
with
$\nu = d/2 -1 $.
In addition,
even in the notoriously difficult
cases of anisotropic singularities~\cite{HEC:dipole_QM-anomaly,HEC:CQManomaly-CQM},
a reduction process can be justified in any number of
dimensions to an effective one-dimensional form similar to 
Eq.~(\ref{eq:singular-QM_radial}). In these  cases, the usual quantum-mechanical interpretation
can be enforced.

As it turns out,
the analysis of physical systems using 
Eq.~(\ref{eq:singular-QM_radial})
is {\em not limited to nonrelativistic quantum mechanics\/}.
In effect,
equations with a leading singularity 
of the form~(\ref{eq:singular-QM_radial})  
arise from a reduction process 
in miscellaneous applications such as the near-horizon physics 
and concomitant thermodynamics of black 
holes~\cite{BH_thermo_CQM,semiclassical_BH,near_horizon,padmanabhan}, 
D-branes~\cite{generic_D-brane_IQP_1,generic_D-brane_IQP_2,generic_D-brane_IQP_3,park_D-brane_IQP},
gauge theories~\cite{HEC:CQManomaly-CQM,QED_gauge},
and quantum 
cosmology~\cite{quantum-cosmology_PIoline-Waldron,quantum-cosmology_Craps-Hertog-Turok}---due 
to space limitations, the list of applications and references is incomplete.
Parenthetically, as mentioned above,
the generic theory still applies but
care should be exercised in these cases when 
 the angular momentum does not appear at the same order 
 with the inverse square potential~\cite{HEC:CQManomaly-CQM,BH_thermo_CQM,semiclassical_BH}.

\subsection{SQM: singularity waves}
\label{sec:singularity-waves}

From the foregoing discussion,
the nature of the singular point $r=0$ in 
Eq.~(\ref{eq:singular-QM_radial}) 
unambiguously leads to the generic definition of {\em singular potentials\/}.
As we will see next, it is possible to derive the explicit form of the solutions 
as $r \rightarrow 0$
 and verify that there is a
breakdown of the boundary condition at the  
singularity  for the sufficiently attractive case~\cite{spector_RMP,Newton,landau:77,perelomov:70,esposito}
with $p=2$ and for all attractive cases with $p>2$.

Specifically,
in the presence of a sufficiently attractive singular potential,
there exists a set of solutions
\begin{equation}
{\mathcal B}_{\rm sing}
=
\biggl\{
u_{+}(r) , u_{-}(r)
\biggr\}
\;  
\label{eq:singularity-basis}
\end{equation}
that behave as local outgoing/ingoing waves 
with respect to the singularity.
Without loss of generality,
the near-singularity 
waves $u_{\pm} (r)$
are properly defined through their 
functional form in the limit $r\sim 0$, which can be completely characterized by 
the WKB method. 
 Even though
this semiclassical technique is usually
regarded as an ``approximation'' or estimate,
WKB becomes ``asymptotically exact'' near a coordinate
singularity (singular point of differential equation)~\cite{spector_RMP,semiclassical_BH}.
In other words,
to any desired degree of accuracy with respect to $r$
  (with the similarity symbol standing for asymptotic approximation hereafter), 
\begin{equation}
u_{\pm} (r)
\stackrel{( r \rightarrow 0)}{\sim}
\frac{1}{ \sqrt{ k_{_{\rm WKB, sing } } (r) } } 
\,
\exp \!
\left[
\pm 
i
\,
\int^{r}_{r_{0}}
dr' \, 
k_{_{\rm WKB, sing } } (r') 
 \right]
\; ,
\label{eq:WKB_waves}
\end{equation}
where 
$k_{_{\rm WKB, sing } } (r)$ 
is the leading part of the WKB local wavenumber
(with respect to $r \rightarrow 0$)
and $r_{0}$ is an arbitrary integration point.
From the dominant near-singularity contribution,
it follows that
\begin{equation}
k_{_{\rm WKB, sing } } (r) 
\stackrel{( r \rightarrow 0)}{\sim}
\begin{cases}
{\displaystyle 
\sqrt{ {\lambda}} \, r^{-p/2} }
& \mathrm{if } \; p > 2  \\
\\
{\displaystyle
\frac{ \Theta }{ r }
}
& \mathrm{if } \; p = 2 
\; .  
 \end{cases} 
\end{equation}

Thus,
by separate direct integration in Eq.~(\ref{eq:WKB_waves})
for the properly and marginally singular cases,
the singularity waves are given by
\begin{eqnarray}
u_{\pm} (r) 
&
\stackrel{( r \rightarrow 0)}{\sim}
&
\frac{ r^{p/4} }{{\lambda}^{1/4}}
\,
\exp \!
\left[
\left.
\mp 2 \, i \, {\lambda}^{1/2}
\,
 \frac{ 
 r^{ - \left(  p /2 - 1 \right) } 
}{
\left( p - 2 \right)}
\right|_{r_{0}}^{r}
\right]
\label{eq:Wannier_waves_p>2_with-r0} 
\\
&
\stackrel{( r \rightarrow 0)}{\sim}
&
\frac{ r^{p/4} }{{\lambda}^{1/4}}
\,
\exp \!
\left[
\mp 2 \, i \, {\lambda}^{1/2}
\,
 \frac{ 
 r^{ - \left(  p /2 - 1 \right) } 
}{
\left( p - 2 \right)}
\right]
\label{eq:Wannier_waves_p>2} 
\end{eqnarray}
and
\begin{equation}
u_{\pm} (r) \stackrel{( r \rightarrow 0)}{\sim}
{\displaystyle
\sqrt{ \frac{ r }{\Theta } }
\,
\exp \!
\left[ 
\pm i \, \Theta
\,
\ln \left( \mu r \right)
\right]
}
\; .
  \label{eq:Wannier_waves_p=2}
\end{equation}
  For $p>2$ in Eq.~(\ref{eq:Wannier_waves_p>2}),
   the integral in the exponent yields an arbitrary additive constant, which is given by the integration point $r_{0}$;
but this is of higher order in the asymptotic expansion with respect to $1/r$.
It should be noticed that,
 by contrast, the integration constant does not
disappear for $p=2$,
where $\mu$ is a  floating inverse length  that arises from the arbitrary integration point, i.e., $\mu = r_{0}^{-1}  $ in
Eq.~(\ref{eq:Wannier_waves_p=2}), which
can also be written  a simple
power-law behavior with imaginary exponent,
\begin{equation}
u_{\pm} (r)
\stackrel{( r \rightarrow 0)}{\sim}
\sqrt{ \frac{ r }{\Theta } }
\;
\left( \, \mu \, r \, \right)^{ \pm  i \, \Theta}
\; .
\label{eq:Wannier_waves_CQM-proper}
\end{equation} 
Moreover, 
the
case $p=2$ includes the Langer correction~\cite{Langer}
 corresponding the critical coupling $\lambda =1/4$.
 Thus,
 the shifted coupling constant 
 \begin{equation}
 \Theta^{2}
 \equiv
\lambda - 1/4 
\end{equation}
 (or its square root) 
becomes the relevant variable in what follows. 
 Furthermore,
 in the particular case of nonrelativistic quantum mechanics (but not for quantum fields in black hole backgrounds)
  the angular momentum
  is merged with the marginally singular $p=2$ term (same order), leading to 
  an effective  interaction coupling. In that case, all the formulas for CQM 
  in this section should involve the replacements 
  $\lambda \rightarrow   \lambda - (l+\nu)^{2} + 1/4$ and, 
  for the critical coupling per angular-momentum channel~\cite{HEC:DTII}:
  $\lambda^{(*)} = (l+\nu)^{2}$,
   i.e., $\Theta_{l}^{2}= \lambda - \lambda^{(*)}$.
   
 Interestingly,
 these expressions can be combined into the single form 
\begin{equation}
u_{\pm} (r)
\stackrel{( r \rightarrow 0)}{\sim}
\frac{ r^{p/4} }{\tilde{\lambda}^{1/4}}
\,
\exp \!
\left[\mp 2 \, i \, \tilde{\lambda}^{1/2}
\,
 \frac{ 
 r^{ - \left(  p /2 - 1 \right) } 
}{
\left( p - 2 \right)}
\right]
\; ,
\label{eq:Wannier_waves}
\end{equation}
by direct integration of the generic case---but keeping the integration point $r_{0}$ as in 
Eq.~(\ref{eq:Wannier_waves_p>2_with-r0}),
which we omit for the sake of simplicity.
Here,
\begin{equation}
\tilde{\lambda} = 
\left\{
\begin{array}{ll}
\lambda & \; {\rm for} \;   p > 2
\\
\lambda - 1/4 = \Theta^{2} &  \; {\rm for} \; p = 2
\;   .
\end{array}
\right.
\end{equation}
The marginally singular case indeed satisfies Eq.~(\ref{eq:Wannier_waves}),
as can be verified by taking the limit $p \rightarrow 2$ of what appears to be a singular expression. 
In this case, using $p = 2+ \epsilon$,
the exponent  in Eq.~(\ref{eq:Wannier_waves}),
before dropping the integration point $r_{0}$,
becomes 
\begin{equation}
y = \mp 2i \hat{\lambda}^{1/2} \mu^{-\epsilon/2} 
\left(
 r^{-\epsilon/2}/\epsilon -  r_{0}^{-\epsilon/2}/\epsilon
 \right)
 \; , 
 \end{equation}
 where $r_{0} = \mu^{-1}$ is an arbitrary integration point defining a floating undetermined inverse length $\mu$,
 and $\tilde{\lambda } =\hat{\lambda } \mu^{-(p-2)}$, with $\hat{\mu}$ dimensionless, as required by dimensional analysis.
In the limit $\epsilon \rightarrow 0$, with $A^{\epsilon} = 1 + \epsilon \ln A + \ldots$,
the exponent indeed becomes 
$
y = \pm i \Theta \ln \left(\mu r \right)
$.
This simple algebra clearly illustrates the peculiar features involved in CQM, $p=2$:
(i) the existence of a critical coupling; (ii) the scale/conformal  invariance (i.e., SO(2,1) symmetry); 
(iii) the ensuing
emergence of an arbitrary scale $\mu$. These issues will be further explored in 
Section~\ref{subsec:CQM}.

An alternative derivation follows from the solutions of differential equation~(\ref{eq:singular-QM_radial}) with $k=0$,
as the term $k^{2}$
becomes negligible when $r \rightarrow 0$ (i.e., it is of higher order in the asymptotic expansion with respect to $1/r$).
For $p>2$, the solution  is of the form
$u \propto \sqrt{r} \, Z_{-1/n}  \left( - 2 \sqrt{\lambda} \, r^{-n/2}/n \right)$, where $n= p-2>0$
and  $Z_{s} (\xi)$ (with $s=-1/n$)
stands for a 
generic Bessel function.
Notice that as $1/r \rightarrow \infty$, the argument 
can be evaluated with the asymptotics of Bessel functions, 
with the Hankel functions $Z=H^{(1,2)}$ having the correct outgoing/incoming behavior,
$H^{(1,2)}_{s} (\xi)
\! 
 \stackrel{(\xi \rightarrow \infty ) }{\sim}
\!
\sqrt{2/\pi \xi}
\,
\exp
\left\{
\pm
i
 \left[ 
\xi 
- s \pi/2
- \pi/4
\right]
\right\}
$. 
As a result,
for $p>2$,
$u_{\pm} \propto r^{1/2} \left[ r^{-n/2} \right]^{-1/2} e^{\mp 2i  \sqrt{\lambda} \, r^{-n/2}/n}$
(factoring out several constants), 
which
 is indeed of the form~(\ref{eq:Wannier_waves_p>2}), with the correct prefactor given by enforcing WKB normalization.
On the other hand,
the $k r \sim 0$ limit for $ p =2$ of Eq.~(\ref{eq:singular-QM_radial})
is dimensionally homogeneous, i.e.,
it is a Cauchy-Euler differential equation~\cite{Nagle-2000},
whence 
Eq.~(\ref{eq:Wannier_waves_CQM-proper})
follows; and yet again, the factor $\mu$ arises from dimensional homogeneity.

In addition,
it should be noticed that the proper WKB amplitude prefactors in
Eqs.~(\ref{eq:WKB_waves})--(\ref{eq:Wannier_waves})
are needed for a correct local definition compatible with 
probability conservation; moreover,
the ``minimalist'' normalization of Eqs.~(\ref{eq:WKB_waves})--(\ref{eq:Wannier_waves})
 simplifies the current-conservation relationships, but other normalizations are also possible.
These issues are further discussed in Secs.~\ref{sec:singularQM_multichannel-Smatrix} 
and
\ref{sec:asymptotic_Smatrix}---thus
 playing a critical role in specific calculations (see Section~\ref{sec:particular_cases} for particular cases).
 A key feature of the solutions $u_{\pm} (r) $
displayed in Eq.~(\ref{eq:Wannier_waves})
is its independence with respect to the
details of the physics at much longer length scales.
Moreover, such near-singularity form of $u_{\pm} (r) $
is valid for all interactions with $ p  \geq 2$, thus covering all
singular cases of physical interest. 
The particular cases $p  = 2$ and $ p  = 4$ will be 
further discussed in this paper.

We can now see that 
the {\em loss of boundary condition\/}~\cite{HEC:DTII,Schwartz}
in Eq.~(\ref{eq:singular-QM_radial}), i.e., the indeterminacy of the solutions,
is a direct consequence of the defining clear-cut dominance of 
the singular interaction over the kinetic part of the Hamiltonian.
Straightforward scaling of
the terms 
of Eq.~(\ref{eq:singular-QM_radial}) 
shows that {\em strong singular behavior\/} occurs when
a potential diverges in terms of the coordinate $r$
at least as
$V(r) 
\stackrel{( r \rightarrow 0)}{\propto}
- r^{-p}$, 
with $ p  \geq 2$ and $r = 0$ 
being the coordinate singularity.
In effect, for $r \sim 0$,
the preponderance of the singular interaction 
drives the states of the system to an ever increasing oscillatory
behavior
with respect to decreasing values of $r$.
As a consequence, the evolution of the system and the
properties of the associated states for $r \rightarrow 0$
are governed in principle by both
states of the basis set~(\ref{eq:singularity-basis}).

The
building blocks
$u_{\pm}(r)$
of Eq.~(\ref{eq:singularity-basis})
play a preferential role as ``singularity probes,''
i.e., they capture the leading  
behavior of the theory when $r \sim 0$.
Thus,
in terms of these
{\em singularity waves\/},
 the general solution to Eq.~(\ref{eq:singular-QM_radial})  
admits the expansion
\begin{equation}
u  (r) 
=
C^{ \mbox{\tiny $ (+)$} } 
u_{+}(r) 
+  
C^{ \mbox{\tiny $ (-)$} } 
u_{-} (r)
\; ,
\label{eq:wave-function_near-origin-basis_generic}
\end{equation}
which we will conveniently rewrite in the form
\begin{equation}
u  (r) \propto 
\Omega 
\, u_{+}(r) +  u_{-} (r)
\; .
\label{eq:wave-function_near-origin-basis}
\end{equation}
In Eq.~(\ref{eq:wave-function_near-origin-basis})
the proportionality symbol 
indicates that a factor
$
C^{ \mbox{\tiny $ (-)$} } 
$
is extracted
and
\begin{equation}
\Omega = 
\frac{
C^{ \mbox{\tiny $ (+)$} } 
}{
C^{ \mbox{\tiny $ (-)$} } 
}
\; 
\label{eq:singularity-parameter}
\end{equation}
can be regarded 
as a ``singularity parameter''
measuring the relative probability amplitudes of 
outgoing (emission) to ingoing (absorption) waves
associated with the neighborhood of $r=0$.
In essence,
a particular choice of 
 $\Omega$ 
 is tantamount to 
specifying an auxiliary ``boundary condition.''
It should be noticed that, for the particular values $|\Omega| = 1$,  
the analysis of Section~\ref{sec:absorption-emission}
 shows that evolution is unitary, i.e., probability-current-conserving 
and thus corresponds to a self-adjoint differential operator: the effective Hamiltonian is 
self-adjoint---an issue that will be further analyzed in the follow-up paper. 
Notice that this amounts to introducing an {\em arbitrary\/} phase 
$\gamma$
associated with an infinite family of self-adjoint extensions, 
and which leads to the relative phase between the two singularity waves, as first 
proposed in Ref.~\cite{cas:50} and further elaborated in Ref.~\cite{perelomov:70}.
In our multichannel S-matrix language,
this amounts to  
\begin{equation}
\Omega = - e^{2 i \gamma} \Longrightarrow
u_{\rm self-adjoint}
\stackrel{( r \rightarrow 0)}{\propto}
\frac{1}{ \sqrt{ k_{_{\rm WKB, sing } } (r) } } 
 \, \sin \left( \int^{r} k_{_{\rm WKB, sing } } (r') dr' + \gamma \right)
\; 
\end{equation}
(with asymptotic proportionality involving a constant $-2i e^{i\gamma}$).
Notice that the arbitrariness of $\gamma$ leads to an infinite set of quantum theories labeled by specific
values of a
self-adjoint extension parameter.

In short,
unlike the case of regular quantum systems: 
both states 
$u_{\pm} (r)$ are in principle on an equal footing;
and
the various linear 
combinations~(\ref{eq:wave-function_near-origin-basis})
may exhibit different degrees of probability loss or gain.
Clearly,
only the use of additional 
information arising from the ultraviolet physics can circumvent 
this peculiar indeterminacy.
The central result of our proposal
is that this information about the singular interaction 
can be completely subsumed in 
a multichannel framework,
as discussed in the next section.

\section{Multichannel framework
for singular quantum mechanics: 
transfer and scattering matrices}
\label{sec:singularQM_multichannel-Smatrix}

\subsection{Singular potentials: bases and indeterminacy}

The indeterminacy posed by
 a coordinate singularity
involves the set
${\mathcal B}_{\rm sing}
=
\biggl\{
u_{+}(r) , u_{-}(r)
\biggr\}
$
of Eq.~(\ref{eq:singularity-basis}),
whose existence suggests 
the following procedure as a treatment
for this ``pathology.''
In this approach,
 the singularity is now regarded 
as a ``hole'' 
${\mathcal P}_{\rm sing}$
to be excluded from the 
relevant domain 
for Eq.~(\ref{eq:singular-QM_radial}),
i.e,
${\mathcal D}_{0}
=
{\mathcal D} - {\mathcal P}_{\rm sing}$,
with outgoing and ingoing waves connecting
the disjoint parts
$ {\mathcal D}_{0} $
and
${\mathcal P}_{\rm sing}$.
Thus,
 ${\mathcal P}_{\rm sing}$
effectively behaves as a ``singularity channel,'' 
describing quantum-mechanical transference of probability to and from
an ultraviolet sector. 
In this view, 
${\mathcal P}_{\rm sing}$
can exhibit a variety of behaviors
according to the short-scale physics---including
cases where neglecting the role played by ${\mathcal P}_{\rm sing}$
may be construed as apparent 
nonunitarity.

In order to implement this idea,
one additional step is needed.
As in regular quantum mechanics, the observables 
can be determined by measurements performed 
via an observer at asymptotic infinity.
One may also view this process as defining
a ``channel'' that connects the
physical domain 
$ {\mathcal D}_{0} $
to asymptotic infinity.
The details of the
 determination of physical observables, leading to
the {\em asymptotic S-matrix\/} $ S_{\rm asymp} $,
will be further developed in the next section.
These details 
rely on two independent solutions $u_{1,2}(r)$
of Eq.~(\ref{eq:singular-QM_radial})
that behave as outgoing/ingoing waves at infinity.
Correspondingly,
the building blocks
$u_{1,2}(r)$
can serve as another set 
\begin{equation}
{\mathcal B}_{\rm asymp}
=
\biggl\{
u_{1}(r) , u_{2}(r)
\biggr\}
\; 
\label{eq:asymptotic-basis}
\end{equation}
of the
 two-dimensional space of solutions, 
distinct from that of Eq.~(\ref{eq:singularity-basis}),
i.e.,
\begin{equation}
u  (r) 
=
C^{ \mbox{\tiny $ (1)$} } 
u_{1}(r) 
+ 
C^{ \mbox{\tiny $ (2)$} } 
  u_{2} (r)
\; ,
\label{eq:wave-function_asymptotic-basis_generic}
\end{equation}
Furthermore,
the solutions to Eq.~(\ref{eq:singular-QM_radial})
turn into a free-wave form $e^{\pm i kr}$
(or quasi-free for the conformal case) as
$r \sim \infty$; correspondingly,
in our treatment,
we will adopt the convention
\begin{equation}
u_{1,2} (r)
\stackrel{( r \rightarrow \infty )}{\sim}
\frac{1}{ \sqrt{k} }
\,
e^{\mp i \pi/4}
\,
e^{\pm i kr} 
\; ,
\label{eq:asymptotic-wave-normalization}
\end{equation}
where the chosen phases will prove convenient for comparison with 
asymptotic expansions of 
Hankel functions, as discussed in Section~\ref{sec:asymptotic_Smatrix}.
It should be noticed that, when the potential has a long-range tail $V (  r )  \sim - \lambda r^{-\delta}$,
the asymptotic behavior is governed by the counterpart of Eq.~(\ref{eq:WKB_waves}) 
at infinity with the extra phase, i.e.,
\begin{eqnarray}
u_{1,2} (r)
& \stackrel{( r \rightarrow \infty)}{\sim} &
e^{\mp i \pi/4}
\,
\frac{1}{ \sqrt{ k_{_{\rm WKB, sing } } (r) } } 
\,
\exp \!
\left[
\pm 
i
\,
\int^{r}
dr' \, 
k_{_{\rm WKB, sing } } (r') 
 \right]
 \label{eq:WKB_waves_infinity}
 \\
& \stackrel{( r \rightarrow \infty)}{\sim} &
 \frac{1}{ \sqrt{k} }
\,
e^{\mp i \pi/4}
\,
e^{\pm i kr} 
\,
e^{\pm i \lambda r^{1-\delta}/2k (1 -\delta)}
\; ,
\label{eq:modified_asymptotic-wave-normalization}
\end{eqnarray}
which yields an asymptotic position-dependent factor $e^{\pm i \lambda r^{1-\delta}/2k (1 -\delta)}$
when $\delta \leq 1$; for the critical case $\delta =1$, considered in~\ref{sec-app:S-matrix_conformally-driven},
the extra factor is 
$e^{\pm i \lambda \ln (\mu r) /2k }$ (with some convenient scale $\mu$).
It should be noticed that, if this procedure looks almost identical 
(Eqs.~(\ref{eq:WKB_waves}) and (\ref{eq:WKB_waves_infinity})), 
it is because the point at infinity is a singular point 
of Eq.~(\ref{eq:singular-QM_radial})---thus, WKB is also asymptotically exact, 
and the  outgoing/incoming waves play a similar role.

\subsection{Multichannel framework}

\begin{figure}[ht]
\centering
\resizebox{4.4in}{!}{\includegraphics{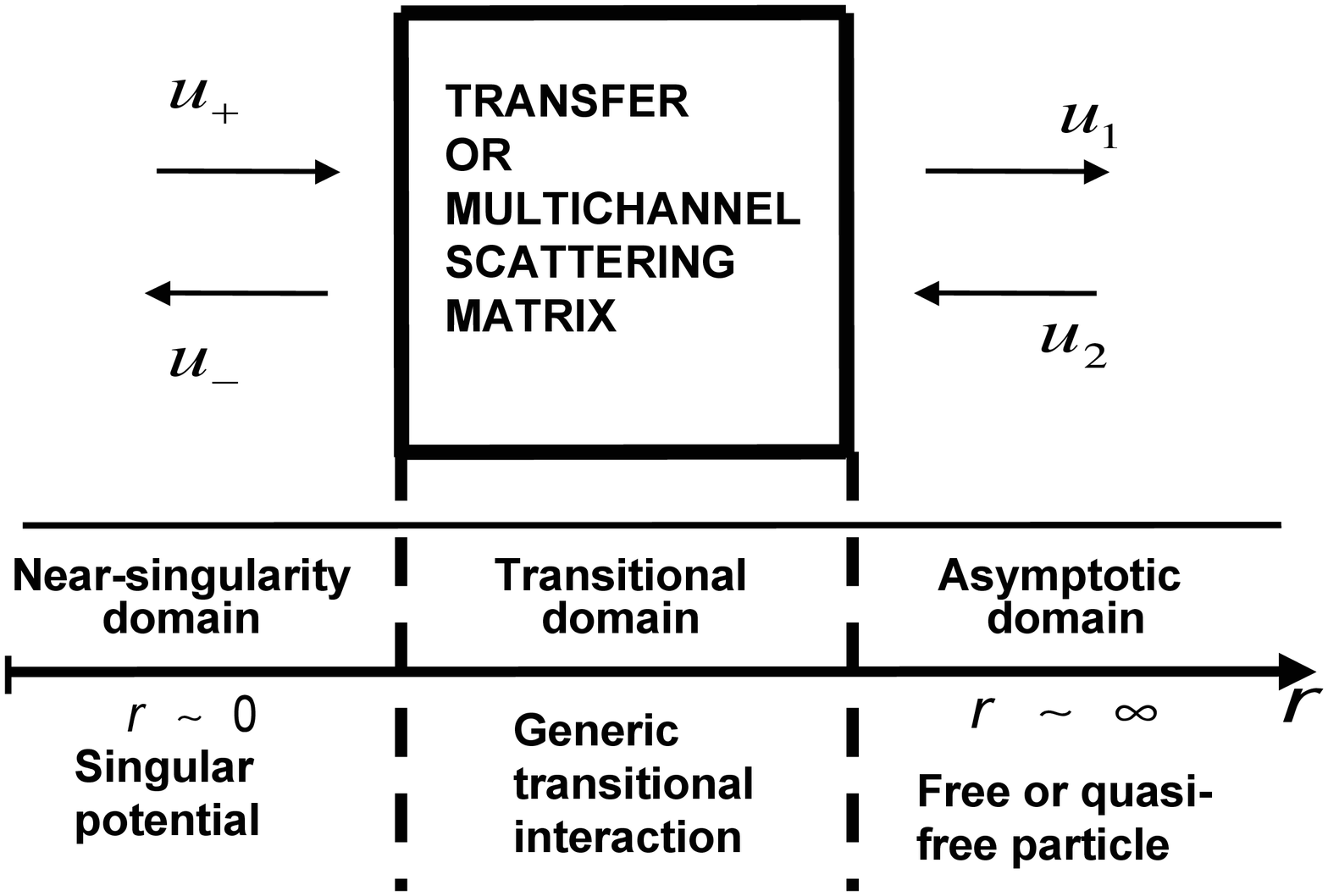}}
\vspace*{-0.5in}
\caption{Graphical depiction of the multichannel framework
as a ``black box'' linking the near-singularity domain
($r \sim 0$)
with the asymptotic domain ($r \sim \infty$).
In all singular systems one could identify at least one relevant scale
$L$ (and possibly more than one)
for the hierarchical definitions $r \ll L$ and $r \gg L$ of
the two domains above. The transitional domain
($r \sim L$) includes a generic interaction
that governs the coefficients of the multichannel matrices.
The arrows to the left and to the right
of the ``black box''
represent the outgoing and ingoing 
 building blocks (waves) at the singularity, 
${\bf u_{+}}$ and ${\bf u_{-}}$,
and at
asymptotic infinity,
${\bf u_{1}}$ and ${\bf u_{2}}$.
}
\label{fig:multichannel}
\end{figure}

 As a consequence of the above choices,
a two-channel framework for one singularity ($N=1$)
is naturally
defined in terms of:
\begin{itemize}
\item
Channel ${\mathcal P}_{\rm sing}$
with set
${\mathcal B}_{\rm sing}$,
Eq.~(\ref{eq:singularity-basis}),
uniquely characterized by the ``ultraviolet'' 
near-singularity behavior~(\ref{eq:WKB_waves})
[or (\ref{eq:Wannier_waves})].
\item
Channel ${\mathcal P}_{\rm asymp}$,
with set
${\mathcal B}_{\rm asymp}$,
Eq.~(\ref{eq:asymptotic-basis}),
uniquely characterized by the infrared asymptotic behavior
(\ref{eq:asymptotic-wave-normalization})
or (\ref{eq:modified_asymptotic-wave-normalization}).
\end{itemize}
The relationship between the basis sets
is determined by the 
physics in the transitional region,
which may include any generic potential,
as displayed in the geometry of
Fig.~\ref{fig:multichannel}.
This connection 
between the singularity,
described via 
$\left[ u^{+} \; \; \; u^{-} \right]$,
and asymptotic
infinity,
described via 
$\left[ u_{1} \; \; \; u_{2} \right]$,
 can be analytically represented in matrix
form
\begin{equation}
u_{j} (r)= 
\sum_{\sigma = \pm}
\alpha_{j}^{\sigma} 
\,
u_{\sigma}(r)
\; \; \; \; \; \; \; \;
({\rm for} \; \; j=1,2)
\; ,
\label{eq:basis-transformation}
\end{equation}
i.e.,
\begin{equation}
\left[
\begin{array}{cc}
u_{1} & u_{2} 
\end{array}
\right]
=
\biggl[
\begin{array}{cc}
u_{+} & u_{-} 
\end{array}
\biggr]
\; \;
\biggl[ \; \;
 \mbox{\boldmath $ \alpha$}
\; \;
\biggr]
\; .
\label{eq:basis-transformation_matrix}
\end{equation}
Thus, for the all-important case of one singular interaction,
the multichannel framework is encapsulated in  
a transfer matrix
\begin{equation}
\biggl[ 
\; \;
 \mbox{\boldmath $ \alpha$} 
\; \;
\biggr]
=
\left[
\begin{array}{cc}
\alpha^{+}_{1}
& 
\alpha^{+}_{2} 
\\
\alpha^{-}_{1}
&
\alpha^{-}_{2}
\end{array}
\right]
\; ,
\label{eq:transfer-matrix_structure0}
\end{equation}
with the elements $\alpha_{j}^{\sigma} $
defined by the transformation Eq.~(\ref{eq:basis-transformation}).
Eq.~(\ref{eq:basis-transformation_matrix})
is equivalent to the relation
\begin{equation}
\left( 
\begin{array}{c}
C^{ \mbox{\tiny $ (+)$} } 
\\ 
C^{ \mbox{\tiny $ (-)$} } 
\end{array}
\right)
=
\biggl[ 
\; \;
 \mbox{\boldmath $ \alpha$} 
\; \;
\biggr]
\,
\left( 
\begin{array}{c}
C^{ \mbox{\tiny $ (1)$} } 
\\ 
C^{ \mbox{\tiny $ (2)$} } 
\end{array}
\right)
\; 
\label{eq:multichannel_transfer-amplitudes}
\end{equation}
between the amplitude coefficients.

The elements of the first column are selected via 
$C^{ \mbox{\tiny $ (1)$} } =1,
C^{ \mbox{\tiny $ (2)$} } = 0 $:
let
$\alpha^{+}_{1} \equiv \alpha$,
$\alpha^{-}_{1} \equiv \beta$,
where $\alpha$ and $\beta$ are arbitrary complex parameters.
Similarly, the elements of the second column correspond to
$C^{ \mbox{\tiny $ (1)$} } =0,
C^{ \mbox{\tiny $ (2)$} } = 1 $;
however, this selection amounts to time-reversal of 
the first column in the form:
$\alpha^{+}_{2}
= \overline{ \beta  }$,
and
$\alpha^{-}_{2}
=\overline{ \alpha  }$,
where the
overbar notation represents the time-reversed variables---i.e., the ``in'' asymptotic state
is the time-reversed ``out'' asymptotic state, and likewise with the near-singularity states.
Then, when the condition of
time-reversal symmetry applies:
$\alpha^{+}_{2}
= \overline{ \beta  }
=  \beta^{*}$
and
$\alpha^{-}_{2}
= \overline{ \alpha  }
=  \alpha^{*}$.
This is known from general properties 
(as a consequence of  Eq.~(\ref{eq:singular-QM_radial})),
but can also be directly rederived by writing
$u_{1} = \alpha u_{+} + \beta u_{-}$
from Eq.~(\ref{eq:basis-transformation}),
followed by complex conjugation
$u_{1}^{*}= \alpha^{*} u_{+}^{*} + \beta^{*} u_{-}^{*}$
and identification of the basis functions 
 $u_{\pm}^{*} = u_{\mp}$
and 
 $u_{1,2}^{*} = u_{2,1}$, whence $u_{2} = \beta^{*} u_{+} + \alpha^{*} u_{-}$.
Then,
the transfer matrix has the generic time-reversal invariant form
\begin{equation}
\left[ 
 \mbox{\boldmath $ \alpha$} 
\right]
=
\left[
\begin{array}{cc}
\alpha \; \; \;  & \beta^{*} 
\\
\beta \; \;  \;  & \alpha^{*}
\end{array}
\right]
\; .
\label{eq:transfer-matrix_structure}
\end{equation}
This restriction reduces 
$
\left[ 
 \mbox{\boldmath $ \alpha$} 
\right]
$
to a dependence from 8 
real parameters to only 
four real parameters (namely, time-reversal amounts to 4 real constraints).
As we will see in the next subsection, 
there is an additional constraint related to the property of current conservation.

\subsection{Currents and associated properties}
\label{sec:currents}

 Eq.~(\ref{eq:singular-QM_radial})
is of Sturm-Liouville type, for which, in general, 
 the Wronskian 
$W \biggl[  u,v \biggr]  $
of two solutions 
$u$ and $v$ is
completely determined by the coefficients of the differential equation~\cite{Nagle-2000}. 
This also applies 
to $W \biggl[  u^{*},v \biggr]  $ due to the reality of the coefficients.
In particular,
for  a given eigenvalue in
Eq.~(\ref{eq:singular-QM_radial}), this implies that 
$W \biggl[  u^{*},v \biggr]  = {\rm const}$;
under such conditions, with $u=v$ and setting $J[u]=W[u^{*},u]/i$,
a conserved current is defined
by the familiar expression
\begin{equation}
J
\biggl[ 
 u 
 \biggr] 
= \frac{1}{i} u^{*} ( r ) u' ( r ) + {\rm c.c.}
\; ,
\label{eq:multichannel-current}
\end{equation}
where ${\rm c.c.} $
stands for the complex conjugate and $u' ( r )=du/dr$.
This quantity is indeed proportional to the quantum-mechanical probability current in ordinary quantum mechanics
(viz.\/, in natural units $\hbar =1 $
and $m=2$).

With the conventional normalizations chosen in Eqs.~(\ref{eq:WKB_waves}) 
and
(\ref{eq:asymptotic-wave-normalization}),
 Eq.~(\ref{eq:multichannel-current}) yields the constant currents
 \begin{equation}
 J 
 \biggl[ 
  u_{\pm} 
  \biggr] 
  =  \pm 2
  \; \; \;
  ,
  \; \; \;
J \biggl[  u_{1,2} \biggr]  =  \pm 2
\;
\label{eq:current_bases}
\end{equation}
  for the given bases;
 and
 \begin{eqnarray}
J 
\biggl[ 
C^{ \mbox{\tiny $ (+)$} } 
u_{+}(r) 
+  
C^{ \mbox{\tiny $ (-)$} } 
u_{-} (r)
  \biggr] 
& = & 
2 \left(  
\left| C^{ \mbox{\tiny $ (+)$} }  \right|^{2} - \left| C^{ \mbox{\tiny $ (-)$} }  \right|^{2}
\right)
\label{eq:Pythagorean_sing-basis}
\\
J 
\biggl[ 
 C^{ \mbox{\tiny $ (1)$} } 
u_{1} ( r )
+  
C^{ \mbox{\tiny $ (2)$} } 
u_{2} ( r )
\biggr]
& = &
2 \left(  
\left| C^{ \mbox{\tiny $ (1)$} }  \right|^{2} - \left| C^{ \mbox{\tiny $ (2)$} }  \right|^{2}
\right)
\label{eq:Pythagorean_asympt-basis}
\; ,
\end{eqnarray}
for the 
generic functions of Eqs.~(\ref{eq:wave-function_near-origin-basis_generic})
and
(\ref{eq:wave-function_asymptotic-basis_generic}).

Eqs.~(\ref{eq:basis-transformation}) and (\ref{eq:basis-transformation_matrix})
give the components of the set~(\ref{eq:asymptotic-basis})
of basis functions
  $u_{1}$ and $u_{2}$ in the singularity basis~(\ref{eq:singularity-basis}),
    in terms of the transfer-matrix coefficients.
From the values of 
$W \biggl[  {u}_{1}^{*}, {u}_{1} \biggr]$,
$W \biggl[  {u}_{2}^{*}, {u}_{2} \biggr]$,
$W \biggl[  {u}_{1}^{*}, {u}_{2} \biggr]$,
and
$W \biggl[  {u}_{1}, {u}_{2} \biggr]$
 (which are all constant for a given eigenvalue $k^2$ in 
 Eq.~(\ref{eq:singular-QM_radial})), 
the following conditions are respectively established: the first two,
\begin{eqnarray}
|\alpha |^{2} - |\beta |^{2}= 1  
 \; \; \; ,  \; \; \;
 |\overline{\alpha} |^{2} - |\overline{\beta} |^{2}= 1  
\label{eq:transfer-matrix-coeff_current-conserv}
\; 
\end{eqnarray}
specify current conservation, i.e.,
Eqs.~(\ref{eq:Pythagorean_sing-basis}) and (\ref{eq:Pythagorean_asympt-basis}), while the last 
 two equations correlate the solutions 
${u}_{1}$ and ${u}_{2}$, which amounts to the basic set of time-reversal relations
\begin{equation}
 \overline{ \beta  }
=  \beta^{*}
\; \; \; \; \; \;
\overline{ \alpha  }
=  \alpha^{*}
\; .
\label{eq:transfer-matrix-coeff_T-reversal}
\end{equation}
The set of ancillary current-conservation
equations~(\ref{eq:transfer-matrix-coeff_current-conserv}), 
which amount to just a single condition under time-reversal invariance,
further restricts the degrees of freedom of the transfer matrix to only 3 real parameters.
 This exhausts all the basic restrictions, because the number of such Wronskians involving 
 the basis~(\ref{eq:asymptotic-basis}) and its complex conjugate is actually 10, but two of these are trivially zero and 
 the other four are equivalent (by complex conjugation) to the 4 
 relations~(\ref{eq:transfer-matrix-coeff_current-conserv})--(\ref{eq:transfer-matrix-coeff_T-reversal}).

It should be noticed that,
in scattering theory, a ``mixed basis'' is usually considered, which essentially consists of the outgoing states on either side, i.e., 
$u_{1}$ and $u_{-}$.
These linearly independent states are usually defined with an ad-hoc normalization 
\begin{eqnarray}
\check{u}_{1} 
& = &
\left\{
\begin{array}{ll}
u_{+} + {\mathcal R} u_{-}
\; , &  \; \; {\rm for} \; x \sim 0
\\
 {\mathcal T} u_{1}
 \; , &  \; \; {\rm for} \; x \sim \infty
\end{array}
\right.
\label{eq:tilde-u_1}
\\
\check{u}_{-}
& = &
\left\{
\begin{array}{ll}
u_{2} + {\mathcal R'} u_{1}
\; , &  \; \; {\rm for} \; x \sim \infty
\\
 {\mathcal T'} u_{-}
 \; , &  \; \; {\rm for} \; x \sim 0
 \; .
\end{array}
\label{eq:tilde-u_-}
\right.
\end{eqnarray} 
From the values of 
$W \biggl[  \check{u}_{1}^{*}, \check{u}_{1} \biggr]$,
$W \biggl[  \check{u}_{-}^{*}, \check{u}_{-} \biggr]$,
$W \biggl[  \check{u}_{1}^{*}, \check{u}_{-} \biggr]$,
and
$W \biggl[  \check{u}_{1}, \check{u}_{-} \biggr]$,
the following conditions are respectively established:
\begin{eqnarray}
 |{\mathcal R}|^{2}  +  |{\mathcal T}|^{2} = 1  \; \; \; ,  \; \; \;
 |{\mathcal R'}|^{2}  +  |{\mathcal T'}|^{2} = 1
\label{eq:S-matrix-coeff_current-conserv}
 \\
{\mathcal R^{*} } {\mathcal T'} + {\mathcal T^{*} } {\mathcal R'} = 0
\; \; \; ,  \; \; \;
{\mathcal T'} = {\mathcal T}
\label{eq:S-matrix-coeff_T-reversal}
\; .
\end{eqnarray}
The first two equations specify current conservation, i.e.,
via Eqs.~(\ref{eq:Pythagorean_sing-basis}) and (\ref{eq:Pythagorean_asympt-basis}); while the
last two correlate the solutions 
$\check{u}_{-}$ and $\check{u}_{1}$, 
in a way that  amounts to the basic set of time-reversal relations~\cite{Stokes-relations},
as implied by the time reversal of  Eq.~(\ref{eq:singular-QM_radial}).
 As for the basis~(\ref{eq:asymptotic-basis})  
  (with transfer-matrix coefficients) above, 
 no additional restrictions are implied by current conservation and time-reversal invariance, 
 because the number of Wronskians involving the basis 
 (\ref{eq:tilde-u_1})--(\ref{eq:tilde-u_-})
 and its complex conjugate is also 10, 
 with two of these being trivial and the other four being equivalent (by complex conjugation) to the 4 relations
(\ref{eq:S-matrix-coeff_current-conserv})--(\ref{eq:S-matrix-coeff_T-reversal}).
 
The parameters in Eqs.~({\ref{eq:tilde-u_1}) and ({\ref{eq:tilde-u_-})
are the familiar {\em amplitude transmission\/}
(${\mathcal T}$,   ${\mathcal T'} $)
{\em    and reflection\/}
(${\mathcal R}$,   ${\mathcal R'} $)
 {\em coefficients\/}, 
 which we have conveniently normalized 
via  
Eqs.~(\ref{eq:Wannier_waves})
and (\ref{eq:asymptotic-wave-normalization}).
Their squared moduli represent the ratio of transmitted or reflected currents relative to the incident currents for the
solutions~(\ref{eq:tilde-u_1}) and ({\ref{eq:tilde-u_-}).
As we will see next, this particular basis set can be compactly described by the S-matrix framework.

\subsection{S-matrix}

The S-matrix $\left[ \; {\bf S} \; \right] $
provides an alternative
representation for the two-channel framework,
with its well-known properties in all areas of physics and applied science.
Moreover, this framework
lends itself for a generalization to any number of channels.
In this 
approach,
the ``in'' 
states 
$\biggl[ 
\begin{array}{cc}
u_{+} \, & \, u_{2} 
\end{array}
\biggr]
$
with respect to the 
transitional domain 
(``black box'' in Fig.~\ref{fig:multichannel}) are 
related to the corresponding ``out'' states
$
\biggl[
\begin{array}{cc}
u_{-} \, & \, u_{1} 
\end{array}
\biggr]
$
by the multichannel S-matrix
\begin{equation}
\left[ 
\; 
{\bf S}
\; 
\right] 
=
\left[
\begin{array}{cc}
S^{-}_{+}
 & 
S^{-}_{2}
\\
S^{1}_{+} 
& 
S^{1}_{2}
\end{array}
\right]
\; ,
\label{eq:multichannel_S-matrix}
\end{equation}
so that 
\begin{equation}
\biggl[
\begin{array}{cc}
u_{+} \, &  \, u_{2} 
\end{array}
\biggr]
=
\biggl[
\begin{array}{cc}
u_{-} \, &  \, u_{1} 
\end{array}
\biggr]
\;
\left[
\; 
{\bf S}
 \; 
\right]
\; .
\label{eq:multichannel_S-matrix_waves}
\end{equation}
Eq.~(\ref{eq:multichannel_S-matrix_waves})
is equivalent to the relation
\begin{equation}
\left( 
\begin{array}{c}
C^{ \mbox{\tiny $ (-)$} } 
\\ 
C^{ \mbox{\tiny $ (1)$} } 
\end{array}
\right)
=
\left[ 
{\bf S}
\right]
\,
\left( 
\begin{array}{c}
C^{ \mbox{\tiny $ (+)$} } 
\\ 
C^{ \mbox{\tiny $ (2)$} } 
\end{array}
\right)
\; 
\label{eq:multichannel_S-matrix_amplitudes}
\end{equation}
between the amplitude coefficients.

The S-matrix 
$\left[ 
{\bf S}
\right]
$
can be rewritten in terms
of the transmission and reflection
coefficients associated with the passage through
the transitional domain displayed in Fig.~\ref{fig:multichannel},
as the following argument shows.
Let 
${\mathcal T}$ and ${\mathcal R}$
be the transmission and reflection coefficients
for propagation from left to right;
and ${\mathcal T}'$ and ${\mathcal R}'$
the transmission and reflection coefficients 
for propagation from right to left.
In the two-step argument represented in
Fig.~\ref{fig:S-matrix_coefficients}:
consider first
an incident right mover with ``initial amplitudes'' 
$
C^{ \mbox{\tiny $ (+)$} } =1
$
and 
$
C^{ \mbox{\tiny $ (2)$} } =0
$,
leading to a transmitted amplitude
 $
C^{ \mbox{\tiny $ (1)$} } 
 = {\mathcal T}$ 
and
a reflected amplitude 
 $
 C^{ \mbox{\tiny $ (-)$} } 
 = {\mathcal R}$,
as shown by the first entries 
in Fig.~\ref{fig:S-matrix_coefficients}.
\begin{figure}[ht]
\centering
\hspace*{-0.4in}
\resizebox{4.4in}{!}{\includegraphics{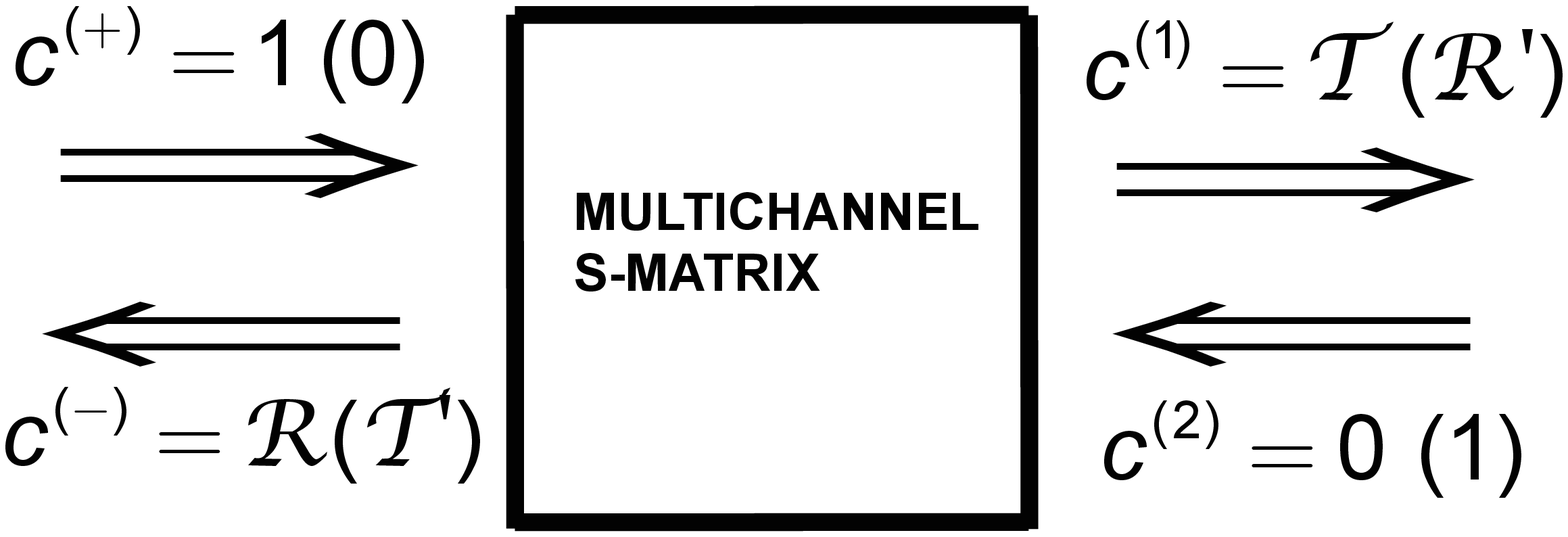}}
\vspace*{-1.2in}

\caption{
Computation of the elements of the multichannel S-matrix
in terms of reflection and transmission coefficients.
The coefficients ${\mathcal T}$ and ${\mathcal R}$ 
are associated with transmission and reflection 
by the ``black box'' or transitional
domain, as caused by the generic interaction,
for right movers---they determine the first column of
the S-matrix. 
Likewise, the coefficients
${\mathcal T}'$ and ${\mathcal R}'$ are
associated with
left movers (in parentheses)---they determine the second column of
the S-matrix.
The double arrows
indicate that the attached labels are amplitude coefficients
(as opposed to the single arrows of Fig.~\ref{fig:multichannel}
representing basis functions).}
\label{fig:S-matrix_coefficients}

\end{figure}
Then, substituting the initial amplitudes 
in Eq.~(\ref{eq:multichannel_S-matrix_amplitudes}),
the first column of 
$\left[ 
{\bf S}
\right]
$
is selected as
$\left( 
\begin{array}{c}
{\mathcal R}
 \\ 
{\mathcal T}
\end{array}
\right)
$.
In a similar manner,
as shown by the second entries 
in Fig.~\ref{fig:S-matrix_coefficients},
for an incident left mover with ``initial amplitudes'' 
$
C^{ \mbox{\tiny $ (2)$} } 
=1$ 
and 
$
C^{ \mbox{\tiny $ (+)$} } 
 = 0$, 
leading to a transmitted amplitude
 $
 C^{ \mbox{\tiny $ (-)$} } 
  = {\mathcal T}'$ 
and
a reflected amplitude 
 $
 C^{ \mbox{\tiny $ (1)$} } 
 = {\mathcal R}'$,
Eq.~(\ref{eq:multichannel_S-matrix_amplitudes})
selects the second column of 
$\left[ 
{\bf S}
\right]
$:
$\left( 
\begin{array}{c}
{\mathcal T}'
 \\ 
{\mathcal R}'
\end{array}
\right)
$.
As a result,
\begin{equation}
\left[ 
{\bf S}
\right]
=
\left[
\begin{array}{cc}
{\mathcal R}
&
{\mathcal T}'
\\
{\mathcal T}
&
{\mathcal R}'
\end{array}
\right]
\; ,
\label{eq:multichannel_S-matrix_coeffs}
\end{equation}
i.e.,
${\bf S}$
is completely
characterized 
in terms
of the transmission coefficients:
${\mathcal T} = S^{1}_{+} $
 and ${\mathcal T}' = S^{-}_{2}$,
and reflection coefficients:
${\mathcal R} = S^{-}_{+} $ 
and ${\mathcal R}' = S^{1}_{2}$,
for the transitions left-right and right-left 
connecting the singularity channel to asymptotic infinity.
This representation
of the S-matrix
is guaranteed to be {\em unitary\/}
due to probability conservation;
and {\em symmetric\/},
whenever time-reversal invariance is satisfied.
These properties are given by 
Eqs.~(\ref{eq:S-matrix-coeff_current-conserv})
and
(\ref{eq:S-matrix-coeff_T-reversal}).

Moreover, the corresponding connection with the
transfer matrix~(\ref{eq:transfer-matrix_structure})
is given by the ratios
\begin{eqnarray}
S^{-}_{+} = {\mathcal R} &  = & 
\frac{\beta }{ \alpha  }
\label{eq:reflection-alpha_connection}
\\
S^{1}_{+} =
{\mathcal T} & = & 
\frac{1 }{ \alpha }
\label{eq:transmission-alpha_connection}
\; .
\end{eqnarray}
As can be easily verified,
Eqs.~(\ref{eq:transfer-matrix-coeff_current-conserv})
and
(\ref{eq:transfer-matrix-coeff_T-reversal})
are thus equivalent to
Eqs.~(\ref{eq:S-matrix-coeff_current-conserv})
and
(\ref{eq:S-matrix-coeff_T-reversal}).

 It should be pointed out
that while the transfer matrix
is specific to our analysis of a single-singularity 
system, the multichannel S-matrix
allows for generalizations to an arbitrary number 
$N$ of singularities
(via an $(N+1) \times (N+1)$ matrix).

\section{Asymptotic S-matrix}
\label{sec:asymptotic_Smatrix}

\subsection{Singular-system asymptotics}

A crucial point in this framework is that
the standard scattering parameters
 measured by an observer at asymptotic infinity
are not simply conveyed by the 
S-matrix~(\ref{eq:multichannel_S-matrix}).
By contrast, the scattering parameters are to be provided
through the usual algorithms of regular quantum mechanics 
by means of an {\em asymptotic S-matrix\/}
$S_{\rm asymp}$, 
which 
is defined for each angular momentum in terms of the 
expansion
\begin{equation}
u (r)
 \stackrel{(r \rightarrow \infty)}{\sim}
 \sqrt{r}
\left[
A^{ \mbox{\tiny $ (1)$} } (k) 
 H^{(1)}_{l+\nu} (kr)
+
A^{ \mbox{\tiny $ (2)$} } (k)
H^{(2)}_{l +\nu} (kr)
\right]
\; ,
\label{eq:asympt_exact_sol}
\end{equation}
with $H^{(1,2)}_{l + \nu} (kr)$ being Hankel functions.
Eq.~(\ref{eq:asympt_exact_sol})
leads directly to the
S-matrix as the simple ratio
\begin{equation}
S_{\rm asymp}
=
\frac{ 
A^{ \mbox{\tiny $ (1)$} } 
}{
A^{ \mbox{\tiny $ (2)$} }
}
\;  ;
\label{eq:scatt_matrix_from_coeff}
\end{equation}
thus, we can conveniently rewrite
\begin{equation}
u (r)
\stackrel{( r \rightarrow \infty)}{\propto}
\sqrt{r} 
\,
\left[
S_{\rm asymp}
\,
H^{(1)}_{l + \nu} (kr) +
 H^{(2)}_{l + \nu} (kr)
\right]
\label{eq:asympt_exact_sol_proportional}
\; .
\end{equation}
Furthermore, 
an {\em asymptotic\/} exponential approximation to
Eq.~(\ref{eq:asympt_exact_sol_proportional}) can be derived from
the identity~\cite{Abramowitz:72}
$H^{(1,2)}_{p} (\xi)
\! 
 \stackrel{(\xi \rightarrow \infty ) }{\sim}
\!
\sqrt{2/\pi \xi}
\,
\exp
\left\{
\pm
i
 \left[ 
\xi 
- p \pi/2
- \pi/4
\right]
\right\}
$,
and  compared against
the $r \sim \infty$
form of the more general
expansion~(\ref{eq:wave-function_asymptotic-basis_generic})
for each singular interaction,
with
Eq.~(\ref{eq:asymptotic-wave-normalization}) 
restricting its normalization properties.
The ensuing comparison of the resolution
\begin{equation}
u (r) 
=
C^{ \mbox{\tiny $ (1)$} } 
u_{1}(r) 
+ 
C^{ \mbox{\tiny $ (2)$} } 
  u_{2} (r)
\propto 
\hat{S}_{\rm asymp} 
\, 
u_{1} (r) +  u_{2} (r) 
\; 
\label{eq:wave-function_asymptotic-basis}
\end{equation}
with Eqs.~(\ref{eq:asympt_exact_sol})--(\ref{eq:asympt_exact_sol_proportional})
leads to
$
A^{ \mbox{\tiny $ (1)$} }/
A^{ \mbox{\tiny $ (2)$} }
=
e^{i \pi \left( l+\nu \right)} 
C^{ \mbox{\tiny $ (1)$} }/
C^{ \mbox{\tiny $ (2)$} }
$, 
i.e.,
it yields the factorization
\begin{equation}
S_{\rm asymp} 
= 
e^{i \pi \left( l+\nu \right)} 
\, 
\hat{S}_{\rm asymp} 
\; ,
\label{eq:S-matrix_and_reduced-S-matrix}
\end{equation}
in terms of 
the {\em reduced matrix elements\/}
$\hat{S}_{\rm asymp}$ 
and an $l$- and $d$-dependent phase factor.

Given
the expansions of Eqs.~(\ref{eq:wave-function_near-origin-basis})
and (\ref{eq:wave-function_asymptotic-basis}),
which amount to two different resolutions
of the wave function,
the ``components'' $\hat{S}_{\rm asymp}$ and $\Omega$ 
are related
via the matrix equation
\begin{equation}
\left[
\begin{array}{c}
\; \Omega \; 
\\
 1
\end{array}
\right]
\propto
\Biggl[
\;  \; 
 \mbox{\boldmath $  \alpha $} 
\; \; 
\Biggr]
\,
\left[
\begin{array}{c}
\! \! 
\hat{S}_{\rm asymp} 
\! \! 
\\
 1
\end{array}
\right]
\; .
\end{equation}
As this is a proportionality relation,
after taking appropriate ratios,
$ \hat{S}_{\rm asymp} $ 
is given by the inverse M\"{o}bius transformation~\cite{Mobius} 
\begin{equation}
\hat{S}_{\rm asymp}
=
\frac{
\alpha^{*}
 \, \Omega 
 -  \beta^{*} }{
  -  \beta \, \Omega +  \alpha  }
\; 
\label{eq:asymptotic-S-matrix_from-transfer-matrix}
\end{equation}
(by inversion of the transfer matrix
$\left[ \alpha_{j}^{\sigma} \right]$).
Alternatively,
from Eqs.~(\ref{eq:reflection-alpha_connection})
and (\ref{eq:transmission-alpha_connection}),
\begin{equation}
\hat{S}_{\rm asymp} 
=
\Delta 
\,
\frac{
  \Omega
-
 {\mathcal R}^{*}
}{
{\mathcal R}
\,
\, \Omega
-
1
}
\label{eq:asymptotic-S-matrix_from-MC-S-matrix}
\;
\end{equation}
where the functional dependence
includes a pure phase  
\begin{equation}
\Delta =  
 - \frac{ {\mathcal T} }{ {\mathcal T}^{*}}
  =
  \frac{ {\mathcal R'} }{ {\mathcal R}^{*}}
  \label{eq:S-matrix_Mobius-phase}
\end{equation}
in addition to the Blaschke factor
$B (\Omega; {\mathcal R} ) =
  \left( \Omega -  {\mathcal R}^{*} \right)/
  \left( {\mathcal R} \, \Omega -  1 \right)
$,
which provides 
a M\"{o}bius transformation of the hyperbolic type~\cite{Mobius}.
In Eq.~(\ref{eq:S-matrix_Mobius-phase}), time-reversal is used explicitly
in the form
$ {\mathcal R'} = \Delta 
 {\mathcal R}^{*}$; in these terms, e.g., 
 \begin{equation}
\hat{S}_{\rm asymp}(\Omega = 0)
=
{\mathcal R'} 
\label{eq:asymptotic-S-matrix_zero-Omega}
\; ,
\end{equation}
 which calibrates 
 the S-matrix for the $\Omega =0$ case
 as the reflection coefficient from the right (for a left mover)---this is the case 
 of ``total absorption''
that is  most commonly considered in the literature
 (see next subsection).
 Thus, in general, 
 the asymptotic S-matrix
is completely characterized via the left- and right-reflection coefficients:
 \begin{equation}
\hat{S}_{\rm asymp} 
=
\frac{
 1-  \Omega/{\mathcal R^{*}} 
}{
 1-  {\mathcal R}  \Omega
}
\,
\hat{S}_{\rm asymp}(\Omega = 0)
 \label{eq:asymptotic-S-matrix_from-zero-Omega}
\; 
\end{equation}

Inspection of 
Eqs.~(\ref{eq:asymptotic-S-matrix_from-transfer-matrix})--(\ref{eq:asymptotic-S-matrix_from-zero-Omega})
suggests a convenient alternative normalization of the singularity and asymptotic waves, such that
\begin{equation}
\tilde{\Omega} = - \Omega \, e^{i \delta_{\mathcal R} }
\; ,
\label{eq:tilde-Omega}
\end{equation}
and
\begin{displaymath}
\tilde{S} = \hat{S}_{\rm asymp} \, e^{-i\delta}
\; ,
\end{displaymath}
where
$\delta 
= 
\delta_{\Delta}
-
\delta_{\mathcal{R} }
$,
with phases defined via
\begin{equation}
\left\{
\begin{array}{ll}
\Delta & = e^{i\delta_{\Delta} }
\\
\mathcal{R} 
&  =
|\mathcal{R} |
e^{i\delta_{\mathcal{R} } }
\end{array}
\right.
\; .
\end{equation}
It should be noticed, by the definition of $\Delta$, that 
$\delta_{\Delta} = 2 \delta_{\mathcal T} + \pi$
(where ${\mathcal T } = |{\mathcal T }| e^{i \delta_{\mathcal T}}$).
With this alternative normalization,
the asymptotic S-matrix can be parametrized by
\begin{equation}
\tilde{S}
=
\frac{
  \tilde{\Omega}
+
|{\mathcal R} |
}{
|{\mathcal R} |
\,
\,\tilde{\Omega}
+
1
}
\; ,
\label{eq:asymptotic-S-matrix_from-MC-S-matrix2}
\end{equation}
whose functional form is depicted in Fig.~\ref{fig:S-matrix}.

\subsection{Absorption, emission, and probability interpretation}
\label{sec:absorption-emission}

At the conceptual level,
it should be noticed that
the multichannel framework permits a {\em physical interpretation\/}
in which information is not lost---even in cases where absorption or
emission are typically ascribed to an effective nonunitary character of the
singular interaction.
The procedure
that leads to the interpretation of
SQM
in terms of absorption and emission
is a generalization of the treatment of
Refs.~\cite{spector_RMP,vogt_wannier}
and is naturally suggested by the associated 
singularity waves of Eq.~(\ref{eq:WKB_waves}).
The phenomenological parameter
$
\Omega  $
in Eq.~(\ref{eq:singularity-parameter})
represents 
the relative probability amplitudes of 
outgoing (emission)
to ingoing (absorption)
waves.
In this scheme,
$\Omega =0 $ 
corresponds to total
absorption and 
$|\Omega| =\infty $ 
to total
emission for a local observer near the singularity.

In this generalized framework, the asymptotic S-matrix is no longer required to be unitary 
and the associated loss of self-adjointness of the Hamiltonian is
consistent with the singular nature of the interaction.
This peculiar behavior should be contrasted with that of regular potentials, for which 
a non-unitary evolution is only possible via a complex potential rather than via singular boundary conditions.
This important property of singular potentials can be understood from
Eqs.~(\ref{eq:Pythagorean_sing-basis}) and (\ref{eq:Pythagorean_asympt-basis}),
which,
by comparison with Eqs.~(\ref{eq:singularity-parameter}),
(\ref{eq:scatt_matrix_from_coeff}), and
(\ref{eq:S-matrix_and_reduced-S-matrix})
 imply that
 $
\left|  \hat{S}_{\rm asymp}  \right|^{2} -1
=
\left[
\left| \Omega \right|^{2} -1
\right]
\,
|C^{(-)}/C^{(2)}|^{2}
$,
leading to
\begin{equation}
{\rm sgn}
\left( J \right)
=
{\rm sgn}
\left[
\left|  \hat{S}_{\rm asymp}  \right|^{2} -1
\right]
=
{\rm sgn}
\left[
\left| \Omega \right|^{2} -1
\right]
\; ,
\label{eq:S-matrix_beta_sign_connection}
\end{equation}
from which 
the following conclusions can be drawn
(see Fig.~\ref{fig:S-matrix}).

\begin{figure}[ht]
\centering
\vspace*{-0.4in}
\hspace*{-0.75in}
\resizebox{5.5in}{!}{\includegraphics{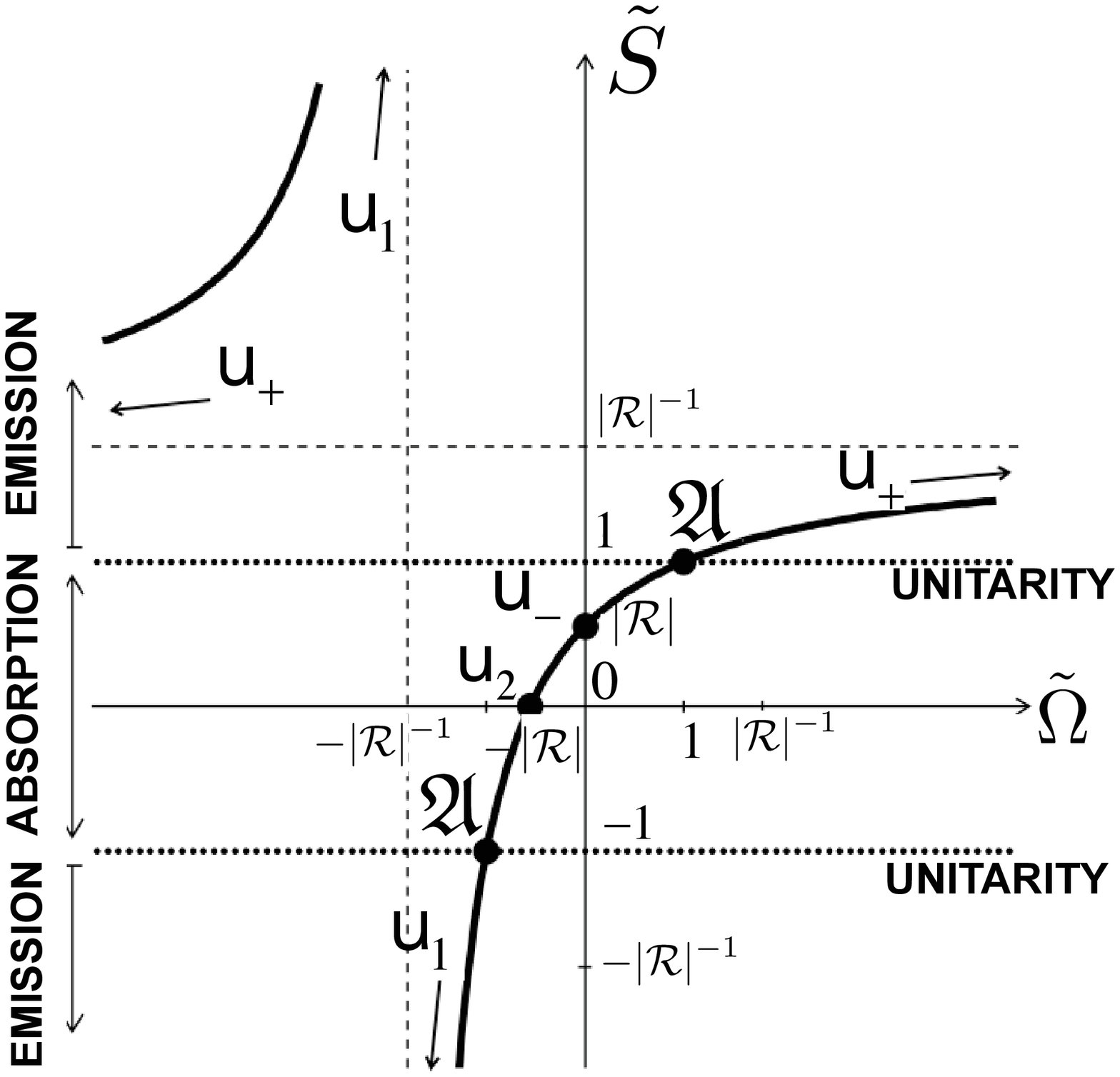}}
\vspace*{-0.2in}

\caption{Asymptotic S-matrix
 of Eqs.~(\ref{eq:asymptotic-S-matrix_from-transfer-matrix})--(\ref{eq:asymptotic-S-matrix_from-zero-Omega}),
 with normalization via
 $\tilde{S} \equiv e^{-i\delta }
  \, \hat{S}_{\rm asymp} $,
as a function of the 
alternative singularity parameter $\tilde{\Omega} = - \Omega e^{i\delta_{\mathcal R} }$
 for generic values of the reflection coefficient's modulus $|\mathcal R|$---notice the phase factor $\Delta$ is
 absorbed in the parameter definitions.
The points singled out in the figure are:
the unitarity points,
$\boldsymbol{\mathfrak{ U}}$,
for self-adjoint states;
 ${\bf u_{_{ \boldsymbol{ \pm}}}}$, 
for the behavior driven by the singularity waves
$u_{\pm} \propto \tilde{u}_{\pm}$ 
of Eqs.~(\ref{eq:WKB_waves})--(\ref{eq:Wannier_waves});
and
 ${\bf u_{1,2}}$
for the behavior driven by the asymptotic waves
$u_{1,2} \propto \tilde{u}_{1,2} $
 of Eq.~(\ref{eq:asymptotic-wave-normalization}).
In addition to the values depicted in this graph, 
the complex-valued S-matrix $\tilde{S}$
and the parameter $\tilde{\Omega}$ 
have arbitrary phases,
so that the points in this figure
can be regarded as real-line intersections 
of mappings 
from the complex $\tilde{\Omega}$ plane 
to the complex $\tilde{S}$ plane;
nonetheless, the points
${\bf u_{\pm}}$ and  ${\bf u_{1,2}}$
do correspond to real $\tilde{\Omega}$,
while the unitarity points correspond to unit circles in both planes.
For conformal quantum mechanics, $\Delta = -Q^{-2i \Theta} = e^{i \delta_{\mathcal R}}
$ 
and $|{\mathcal R}| = e^{-\pi \Theta}$.
}
\label{fig:S-matrix}

\end{figure}

First, we conclude that the net  flux or current $J$ defined from Eq.~(\ref{eq:multichannel-current})
satisfies the conditions:
\begin{itemize}
\item
$J=0$ iff $| \Omega | =1$
\item
$J<0$, i.e.,
the flux is ingoing, 
 iff $|\Omega|<1$,
 with maximum magnitude for $\Omega =0$.
\item
$J>0$, i.e.,
the flux is outgoing,
iff $|\Omega|>1$,
with maximum magnitude for $\Omega =\infty$.
\end{itemize}

Second, 
the various cases can be interpreted and classified in terms of the asymptotic S-matrix as follows:
\begin{enumerate}
\item
Elastic scattering or ``self-adjoint behavior'' is 
defined to correspond to probability conservation,
which is equivalent to the unitarity of the S-matrix,
so that $ \left| \hat{S}_{\rm asymp}   \right| =1$.
The associated  phenomenological
characterization, from Eq.~(\ref{eq:S-matrix_beta_sign_connection}), involves
$|\Omega| =1$
(cf.\ 
Refs.~\cite{perelomov:70,cas:50},
with $\Omega = - e^{2 i \gamma}$ and $\gamma $ being a real phase).

\item
Net local absorption at the singularity amounts to net probability loss,
which is equivalent to the existence of a nonunitary  S-matrix with
elements 
$\left| \hat{S}_{\rm asymp}   \right| <1$.
The associated  phenomenological
characterization, from Eq.~(\ref{eq:S-matrix_beta_sign_connection}), involves
$|\Omega| < 1$.
\item
Similarly, net emission at the singularity 
amounts to net probability gain,
and nonunitary  S-matrix 
elements $\left| \hat{S}_{\rm asymp}   \right| >1$; correspondingly,
$|\Omega| > 1$.
\end{enumerate}

The analysis of this subsection 
is especially relevant for applications where an absorption or capture cross section needs to be computed.
For each channel, the absorptivity is defined by
\begin{equation}
\mathcal{A}
 = 1 - \left|  {S}_{\rm asymp}  \right|^{2} =
1 - \left|  \hat{S}_{\rm asymp}  \right|^{2}
\; .
\label{eq:absorptivity}
\end{equation}
This corresponds to a scattering setup 
 with an ad hoc normalization unity for an incoming asymptotic wave, yielding an
outgoing asymptotic wave of amplitude $ {S}_{\rm asymp}  $. From
Eqs.~(\ref{eq:Pythagorean_sing-basis}) and (\ref{eq:Pythagorean_asympt-basis}),
the conserved current yields a net ingoing flux $\mathcal{A}$ that measures the absorbed probability
(as a fraction of unity).
This also implies that, 
for the particular case with $\Omega =0$, the absorptivity is measured by the transmission coefficient alone:
$\left. \mathcal{A}\right|_{\Omega = 0} = | {\mathcal T}|^{2}$.

In summary, the flux argument of the previous paragraph is  completely general,
leading to a practical rule
for the 
absorption cross section $\sigma_{\rm abs}$;
for example, in $d=3$ dimensions, the usual quantum-mechanical 
partial-$l$ capture cross section is 
$\sigma_{{\rm abs}, l}=  \mathcal{A}_{l} \, \pi k^{-2}$
with 
$\sigma_{{\rm abs},l}= 
\mathcal{A}_{l}
\,
[\Gamma(d)/2]
 \,
 \left[
\Omega_{d-2}/ (d-1)\right] k^{-(d-1)} 
 $
  in $d$ dimensions 
  (according to the volume 
  $\Omega_{d-2}/ (d-1)$
  of the ball bounded by the unit sphere $S^{d-2}$ of $(d-2)$ dimensions)---these 
  values can be easily verified by direct computation with Gegenbauer polynomials~\cite{HEC:DTII}.  
What these general arguments show is that
the ``absorbed flux'' is transferred to another channel
at each singularity point. In short,
this is  how unitarity is restored for singular 
potentials within the multichannel framework.

\section{Particular cases and applications of singular interactions}
\label{sec:particular_cases}

The basic strategy to completely characterize the observables is to find the coefficients of the S-matrix. 
For the case of one singular point outlined in the previous sections, this is systematically achieved via the two-channel framework.
The most efficient algorithm consists in finding the connection between the bases of the two channels via the transfer matrix coefficients.
This involves either finding the resolution of 
Eqs.~(\ref{eq:basis-transformation}) and (\ref{eq:basis-transformation_matrix}) or their inverse transformation.
For the former,
under the conditions of unitarity and time-reversal symmetry, the transfer matrix 
coefficients~(\ref{eq:transfer-matrix_structure})
can be directly obtained from
\begin{equation}
u_{1} = \alpha u_{+} + \beta u_{-}
\; .
\label{eq:u_1from-u_pm}
\end{equation}
For the latter, 
as follows from the inverse transfer matrix,
\begin{equation}
u_{+} = \alpha^{*} u_{1} - \beta u_{2}
\;  .
\label{eq:u_+from-u_12}
\end{equation}
From Eqs.~(\ref{eq:u_1from-u_pm}) and (\ref{eq:u_+from-u_12})
one could derive the other two equations by  
 consistency with the other coefficients via  
$u_{2}= u_{1}^{*}$ and $u_{-}= u_{+}^{*}$; but in practice, only one equation, either
(\ref{eq:u_1from-u_pm}) or (\ref{eq:u_+from-u_12}) suffices for a complete determination of the observables.

\subsection{Conformal quantum mechanics}
\label{subsec:CQM}

Conformal quantum mechanics (CQM)
is selected by the exponent  $ p = 2$
in Eqs.~(\ref{eq:singular-QM_radial})--(\ref{eq:power-law_potential}),
where the interaction is called the inverse square potential (ISP), and
for which the system is scale and conformally 
invariant~\cite{HEC:ISP_letter,HEC:DTI,HEC:DTII,HEC:dipole_QM-anomaly,HEC:CQManomaly-CQM}---with 
an SO(2,1) Schr\"{o}dinger-type symmetry associated with an additional time 
parameter~\cite{Alfaro_Fubini_Furlan:76,Jackiw:ISP,Jackiw_SO(21)}.
The corresponding
 singularity waves involve a logarithmic phase in the exponent
 as described by
Eq.~(\ref{eq:Wannier_waves_p=2}),
or its alternative
power-law expression with imaginary exponent,
Eq.~(\ref{eq:Wannier_waves_CQM-proper}).
From the analysis of Section~\ref{sec:singularity-waves},
these functions require
an arbitrary inverse-length scale $\mu$,
which is also mandatory by general arguments based on dimensional
analysis and scale invariance.
Eq.~(\ref{eq:Wannier_waves_CQM-proper}),
as derived directly from a
Cauchy-Euler differential equation~\cite{Nagle-2000},
highlights the fact that,
sufficiently close to the singularity,
the equation  is not only conformal but also homogeneous. In other words,
all physical scales disappear altogether---this  
scale-invariant singular behavior is the signature of CQM and inevitably yields the arbitrary scale $\mu$.
The robustness of this behavior is specifically 
verified by the generalizations discussed in~\ref{sec-app:S-matrix_conformally-driven}.

 While the $r \sim 0$ behavior of the singularity waves given by Eqs.~(\ref{eq:Wannier_waves_p=2})
and (\ref{eq:Wannier_waves_CQM-proper})
is valid for any system in which the near-singularity physics is conformal,
it is of special interest to examine more closely the {\em global
solutions for the
intrinsically conformal problem\/},
i.e., 
 CQM deprived of any additional long-distance scales.
In this ``pure CQM''  case, the exact solutions 
are given in terms of Bessel functions, i.e.,
\begin{equation}
u_{\pm} (r)
=
N_{\pm} (k; \mu )
\, \sqrt{r} J_{\pm i \Theta } (k r)
\; .
\label{eq:Wannier_waves_pure-CQM}
\end{equation}
In Eq.~(\ref{eq:Wannier_waves_pure-CQM})
 the normalization factor $N_{\pm} (k; \mu) $ is adjusted to 
satisfy Eq.~(\ref{eq:Wannier_waves_CQM-proper})
from the expansion~\cite{Abramowitz:72}
$J_{p}(z) \stackrel{(z \rightarrow 0)}{\sim}  [ \Gamma(p+1) ]^{-1} (z/2)^{p}$,
i.e.,
\begin{equation}
N_{\pm} (k; \mu)
=
\frac{ \Gamma ( 1 \pm i \Theta ) }{ \sqrt{\Theta} }
\,
\left( \frac{k}{2 \mu}  \right)^{ \mp  i \, \Theta}
=
\sqrt{ \frac{ \pi }{ \sinh \left(  \pi \Theta \right) } }
\,
Q^{\mp i \Theta }
\; ,
\label{eq:Wannier_waves_pure-CQM_normalization}
\end{equation}
where
\begin{equation}
Q = \frac{  k 
 }{ 2 \mu  \, e^{- \gamma_{\Theta} } }
\;  ,
\end{equation}
with
$\Gamma ( 1 \pm i \Theta ) = \left| \Gamma ( 1 \pm i \Theta ) \right| \, e^{\pm i \delta_{\Theta}}$
defining
\begin{equation}
\delta_{\Theta}
=
{\rm phase}
\left[
\Gamma (1 +i\Theta )
\right]
\; \; \;  
,
\; \; \;  
\gamma_{\Theta} = - \delta_{\Theta}/\Theta 
\; ,
\end{equation}
and
$ \left| \Gamma ( 1 \pm i \Theta ) \right|
= \sqrt{ \left( \pi \Theta \right)/\sinh \left( \pi \Theta \right)}$
because~\cite{Abramowitz:72} 
$  \Gamma ( 1 +  i \Theta ) \,  \Gamma ( 1 - i \Theta ) = \pi \Theta/\sinh \left( \pi \Theta \right) $.
The required relation between the sets 
${\mathcal B}_{\rm sing}$ 
(given by
Eqs.~(\ref{eq:Wannier_waves_pure-CQM})--(\ref{eq:Wannier_waves_pure-CQM_normalization}))
and
${\mathcal B}_{\rm asymp}$
(given by Eq.~(\ref{eq:asymptotic-wave-normalization}))
can be easily established 
after rewriting the asymptotic conformal waves as
\begin{equation}
u_{1,2}(r)
= e^{\mp \pi \Theta/2} \, \sqrt{\frac{\pi \, r }{ 2} }  \, H^{(1,2)}_{i \Theta } (kr)
\; 
\end{equation}
(from Eq.~(\ref{eq:asymptotic-wave-normalization}) and the asymptotics of Hankel functions)
and applying the Bessel-function identities 
$H_{s}^{(1,2)}(\xi)
= \pm i [ e^{\mp i \pi s} J_{s} (\xi) - J_{-s} (\xi)]/\sin (\pi s)$,
which imply
the exact linear transformations
\begin{equation}
u_{1,2} (r) 
=
\pm
\sqrt{ \frac{\pi}{2} } \,
\frac{1}{ \sinh  \left( \pi \Theta \right) }
\,
\left[
\frac{1}{N_{+}}
e^{\pm \pi \Theta/2}
u_{+}(r)
-
\frac{1}{N_{-}}
e^{\mp \pi \Theta/2}
u_{-}(r)
\right]
\; .
\end{equation}
This conforms with Eq.~(\ref{eq:u_1from-u_pm}) (and its complex conjugate).
As a result, 
the transformation 
coefficients in Eqs.~(\ref{eq:basis-transformation})
and (\ref{eq:basis-transformation_matrix})
become
\begin{eqnarray}
\left\{
\begin{array}{c}
\alpha
\\
\beta
\end{array}
\right\}
\equiv 
 \alpha_{1}^{\pm}
& = &
\pm  \frac{ 1  }{  \sqrt{ 2 \pi \Theta  } }
\,
\Gamma (1 \mp i \Theta)
\,
\, e^{\pm \pi \Theta/2}
\,
\left( \frac{ k  }{2 \mu} \right)^{\pm i \Theta }
\label{eq:transfer-matrix_CQM1}
\\
& = &
\pm  \sqrt{ \frac{ 1  }{ 2 \sinh (\pi \Theta) } }
\, e^{\pm \pi \Theta/2}
\,
Q^{\pm i \Theta }
\; .
\label{eq:transfer-matrix_CQM2}
\end{eqnarray}
Correspondingly,
 from Eqs.~(\ref{eq:S-matrix-coeff_T-reversal}) 
and
(\ref{eq:multichannel_S-matrix_coeffs})--(\ref{eq:transmission-alpha_connection}),
the multichannel S-matrix
is given by
\begin{equation}
\left[ 
\;
{\bf S}
\; 
\right] 
=
\left[
\begin{array}{cc}
- e^{-\pi \Theta} \, Q^{-2i\Theta} 
 & 
\sqrt{ 1 - e^{-2\pi \Theta} } \,  Q^{-i\Theta} 
\\
\sqrt{ 1 - e^{-2\pi \Theta} } \,  Q^{-i\Theta} 
& 
e^{-\pi \Theta}
\end{array}
\right]
\; ,
\label{eq:multichannel_S-matrix_CQM}
\end{equation}
and the asymptotic S-matrix
takes the  generic form 
\begin{eqnarray}
\! \! \! \! \! \! \! \! \! \! \! \! \! \! \!
\hat{S}_{\rm asymp}
& = &
\frac{  
  \Gamma ( 1 + i \Theta ) 
  \,
 e^{\pi \Theta/2}  
\, \left(k/2\mu \right)^{-i \Theta} 
\,
\Omega  
+ 
  \Gamma ( 1 - i \Theta ) 
\, e^{-\pi \Theta/2} 
\left(k/2\mu \right)^{i \Theta} 
}{
  \Gamma ( 1 + i \Theta ) 
\,
e^{-\pi \Theta/2} 
\,
\left(k/2\mu \right)^{-i \Theta} 
\,
 \Omega  
+
  \Gamma ( 1 - i \Theta ) 
  \,
e^{\pi \Theta/2}   
\,
\left(k/2\mu \right)^{i \Theta} 
}
\\
& = &
\frac{  
e^{\pi \Theta/2}  
\,
Q^{-i \Theta} 
\,
\Omega  
+ 
e^{-\pi \Theta/2} 
\,
Q^{i \Theta} 
}{
e^{-\pi \Theta/2} 
\,
Q^{-i \Theta} 
\,
 \Omega  
+
e^{\pi \Theta/2}   
\,
Q^{i \Theta} 
}
\; .
\label{eq:S-matrix_singular-conformal}
\end{eqnarray}
Notice that, for
conformal quantum mechanics,
$|{\mathcal R}| = e^{-\pi \Theta}$,
and
 $\Delta = -Q^{-2i \Theta} = e^{i \delta_{\mathcal R}}
$,
so that 
$\delta_{\mathcal R} =  \delta_{\Delta} $ and $\delta =0$,
leading to $\tilde{\Omega} = - \Omega \Delta$
in Eq.~(\ref{eq:tilde-Omega}).

Moreover,
the scale invariance of CQM leads to the presence of the
arbitrary scale $\mu$ 
in the above expressions 
(manifested as a logarithmic 
phase). As a consequence, 
{\em there exists a factor ambiguity in the elements of the transfer matrix
and the multichannel S-matrix
associated with the conformal singularity.\/}
Specifically,
under the scale change
$\mu \rightarrow \tilde{\mu} $,
the corresponding change in the conformal singularity waves is given by
\begin{equation}
{u}_{\pm}(r; \tilde{ \mu} )
=
\left( 
\frac{ \tilde{ \mu} }{\mu}  
\right)^{ \pm  i \, \Theta}
\,
u_{\pm} (r; \mu)
\;  .
\end{equation}
Then, the transformed transfer matrix 
involves
$\tilde{ \alpha}^{\pm}_{1} 
=
\alpha^{\pm}_{1} 
(\tilde{\mu}/\mu)^{\mp i \Theta}
$, i.e.,
\begin{equation}
\left[ 
\tilde{
 \mbox{\boldmath $ \alpha$}
} 
\right]
=
\left[
\begin{array}{cc}
\left(  \tilde{\mu}/\mu \right)^{ - i \, \Theta}
\, \alpha 
\; \; \;
& 
\; \; \;
\left(  \tilde{\mu}/\mu \right)^{ - i \, \Theta}
\, \beta^{*}
\\
\left(  \tilde{\mu}/\mu \right)^{ i \, \Theta}
\, \beta 
\; \; \;
& 
\; \; \;
\left(  \tilde{\mu}/\mu \right)^{ i \, \Theta}
\, \alpha^{*}
\end{array}
\right]
\; ,
\label{eq:transfer-matrix_structure_rescaled}
\end{equation}
where, as before,
 $\alpha^{+}_{1}  = \alpha$ and $ \alpha^{-}_{1} = \beta$.
In addition, the ``relative components'' of the singularity waves
imply that 
the singularity parameter $\Omega$ satisfies the relation
\begin{equation}
\tilde{\Omega}
 = 
\left( \frac{ \tilde{\mu} }{\mu} 
 \right)^{ - 2 i \, \Theta}
\, \Omega
\; ;
\end{equation}
therefore,
$\Omega$ is only defined up to a 
phase factor---in such a way that
the combination
$
\tilde{ { \Omega } }
\,
\bigl(  \tilde{ {\mu} }  \bigr)^{2 i \, \Theta}
=
\Omega 
\, 
 \mu^{ 2 i \, \Theta}
$
is an invariant under the 
dimensional rescalings
$\mu \rightarrow \tilde{\mu} $.
Finally,
the {\em asymptotic S-matrix\/}~(\ref{eq:S-matrix_singular-conformal})
 is guaranteed to satisfy the 
invariance condition
\begin{equation}
\hat{S}_{\rm asymp}
\biggl[ 
\tilde{\Omega} ;
\left[ 
\tilde{ \mbox{\boldmath $ \alpha$} }
\right]  
\biggr]
=
\hat{S}_{\rm asymp}
\biggl[
\Omega;
\left[ 
 \mbox{\boldmath $ \alpha$} 
\right]
\biggr]
\; ,
\label{eq:asymptotic-S-matrix-rescaled}
\end{equation}
as required by the fixed normalization condition~(\ref{eq:asymptotic-wave-normalization}).

A byproduct  of the arbitrariness with respect to $\mu$ is the
freedom to select any 
normalization of choice.
For example, a naturally convenient selection
would be provided by a  {\em real, energy-independent normalization\/}
of the exact solutions~(\ref{eq:Wannier_waves_pure-CQM}); 
it follows that this would amount to 
\begin{equation}
\hat{N}_{\pm}
=
\frac{ \left| \Gamma ( 1 \pm i \Theta ) \right| }{ \sqrt{\Theta} }
=
\sqrt{ \frac{ \pi }{ \sinh \left(  \pi \Theta \right) } }
\; ,
\end{equation}
which corresponds to $\mu = k e^{\gamma_{\Theta}}/2$,
so that $Q=1$.
If we denote these Bessel-like building blocks by 
\begin{equation}
J_{\pm}(r)
=
\hat{N}_{\pm} 
\, \sqrt{r} J_{\pm i \Theta } (k r)
\; ,
\label{eq:Wannier-Bessel_waves_pure-CQM}
\end{equation}
the relationship between the two bases is given by
\begin{equation}
J_{\pm}(r)
=
\left( \frac{k}{2 \mu}  \right)^{ \pm  i \, \Theta}
\, 
e^{\mp i \delta_{\Theta} }
\,
u_{\pm} (r)
\; ,
\label{near-singularity-bases_connections}
\end{equation}
which leads to
\begin{equation}
\hat{S}_{\rm asymp}
=
\frac{  
e^{\pi \Theta/2}  
\,
\tilde{\Omega}  
+ 
e^{-\pi \Theta/2} 
}{ 
e^{-\pi \Theta/2} 
\,
\tilde{\Omega}  
+
e^{\pi \Theta/2}   
}
=
\frac{  
e^{\pi \Theta}  
\,
\tilde{\Omega}  
+ 
1
}{ 
e^{-\pi \Theta} 
\,
\tilde{\Omega}  
+
1
}
\;
e^{- \pi \Theta}
\; ,
\label{eq:S-matrix_singular-conformal-mod}
\end{equation}
where $ \tilde{\Omega}  =  \Omega \, Q^{-2i\Theta}  $ is the related singularity parameter.
It should be noticed that this equivalent to Eq.~(\ref{eq:asymptotic-S-matrix_from-MC-S-matrix2}).

In summary,
the critical condition that the theory satisfies is that,
regardless of the choice of normalization,
all physical observables are uniquely defined.
However,
the expressions for the multichannel matrices (transfer and S-matrix)
 still exhibit the scale ambiguity inherent in a conformal theory. 
While the form of Eq.~(\ref{eq:S-matrix_singular-conformal})
is useful for a comparison of all known approaches to CQM,
that of Eq.~(\ref{eq:S-matrix_singular-conformal-mod})
suffices for many purposes---i.e,
the quantities 
$\Omega$
and
$\tilde{\Omega}$
only differ by a pure phase factor.

It should be noticed that the definition of asymptotic states
{\em under strict conformal invariance\/}
(i.e.,  in the absence of additional scales)
 is usually not regarded as a
well-posed problem~\cite{Aharony-Maldacena_largeN:2000}.
In our framework,
this is manifested by the long-range behavior of the 
 conformal potential, which inevitably mixes with the free-wave solutions, i.e.,
with the angular momentum,
through the combination 
$\Theta =\sqrt{ \lambda - (l+ \nu)^{2}}$.
However, one could still define the phase shifts by generalizing the approach
used for regular potentials.
The ultimate justification of 
this ad hoc procedure lies in the use of a cutoff scale
beyond which the interaction behaves as a short-range potential
 and the separation
is uniquely defined. 
For example,
via the regularized potential
$\tilde{V}(r) = V(r) f(r)$,
with 
$f(r)=o(1)$ as $r\rightarrow \infty$, 
the scattering matrix and phase shifts are uniquely determined.

The difficulties associated with the conformal potential have been discussed using a variety of 
methods~\cite{gup:93,HEC:ISP_letter,HEC:DTI,HEC:DTII,HEC:dipole_QM-anomaly,HEC:CQManomaly-CQM,beane:00}, which,
remarkably, are subsumed by the multichannel framework. This can be verified by a straightforward analysis
of appropriate limits of Eq.~(\ref{eq:S-matrix_singular-conformal}).
The unitary solutions correspond to the particular case
$|\Omega|= 1$ and can be parametrized as in Ref.~\cite{HEC-SA}, where an
S-matrix technique similar to that of
our present paper is used (restricted to the unitarity sector);
the corresponding solutions in this family coincide with the self-adjoint extensions.
In particular, the different methods, including renormalization approaches, were compared in Ref.~\cite{HEC-SA}
via an analysis of the poles of the S-matrix.
For the unitary solutions,
the physical applications involve miscellaneous realizations and models, 
including the three-body Efimov effect~\cite{Efimov_effect},
dipole-bound anions in molecular physics~\cite{HEC:dipole_QM-anomaly},
QED$_{D}$ (in $D=d+1$ spacetime dimensions)
with
chiral symmetry breaking for strong coupling~\cite{QED_gauge},
black hole thermodynamics~\cite{BH_thermo_CQM,semiclassical_BH,near_horizon,padmanabhan}, 
and the family of Calogero models~\cite{Calogero}.
These examples illustrate some of the  most interesting realizations  of conformal quantum mechanics,
for which the renormalization procedure yields
an anomaly or quantum symmetry breaking in the strong-coupling 
regime~\cite{HEC:CQManomaly-CQM,HEC:anomaly_qm_prelim,Esteve:anomaly}.

\subsection{Inverse quartic potential}
\label{subsec:IQP}

For the inverse quartic potential~\cite{vogt_wannier,IQP_Spector,IQP_Aly-Muller,IQP_Aly-Muller_2,IQP_other},
i.e.,
$V(r) = - \lambda/r^{4}$, the
corresponding Eq.~(\ref{eq:singular-QM_radial})
with
$ p = 4$
can be recast, 
via the 
exponential substitution $\mu r =  e^{\zeta}$ 
of~\ref{sec:exponential-Liouville},
into the canonical form of the  modified Mathieu  
equation (i.e., the Mathieu equation 
of imaginary argument)~\cite{Abramowitz:72,NIST,Gradshteyn-Ryzhik,Meixner-Schafke,Mathieu_McLachlan}
\begin{equation}
\frac{d^2 w}{d\zeta^2} 
-
\left( 
a^2
-
2 h^{2} \cosh 2 \zeta
\right)
w
=
0
\; ,
\label{eq:Mathiew_modified}
\end{equation}
which involves  two dimensionless parameters
\begin{equation}
a =  l + \overline{\nu}
\; \; \; 
; 
\; \; \; 
h^{2} =k \sqrt{\lambda}
\; ,
\end{equation}
where
$\overline{\nu} =d/2 -1$ (the overbar is used
in this subsection  to distinguish this dimensionality parameter from the Floquet
parameter $\nu$ below).
In addition, the characteristic length   
 $\mu^{-1}= \sqrt{\lambda}/h= \lambda^{1/4}/\sqrt{k}$ is
used to repackage the exponentials into the symmetric $\cosh$ form of the 
modified Mathieu equation~\cite{vogt_wannier}.

There are several sets of solution functions and parametrizations for Eq.~(\ref{eq:Mathiew_modified}).
This proliferation reflects in part
its nontrivial nature applied to a wide variety of 
realizations~\cite{Mathieu_McLachlan,Mathieu_applications}---partial comparative summaries
can be found in Refs.~\cite{Abramowitz:72,NIST}.
We focus below on the features and definitions specifically relevant to our scattering problem,
with the notation above in accordance to
Refs.~\cite{park_D-brane_IQP,Muller-Kirsten_QM-book}.

The most widely used set of solutions involves the modified version of the Floquet-type 
Mathieu cosine, sine, and exponential functions:
$Ce_{\nu} (\zeta, h^2)$,
$Se_{\nu} (\zeta, h^2)$,
and
$Me_{\nu} (\zeta, h^2)$,
along with 
the second solutions 
$Fe_{\nu} (\zeta, h^2)$
and 
$Ge_{\nu} (\zeta, h^2)$~\cite{Abramowitz:72,NIST,Gradshteyn-Ryzhik,Meixner-Schafke,Mathieu_McLachlan},
 where $\nu = \nu (a,h) $ is the Floquet parameter or characteristic exponent 
 (associated with the periodicity properties of the ordinary Mathieu equation).

The other commonly used set of solutions, the Mathieu-Bessel functions
$M_{\nu}^{(j)} (\zeta,h)$ ($j=1,2,3,4$) 
are defined by their asymptotic behavior for
${\rm Re} ( \zeta ) \sim \infty$, so that they asymptotically match the corresponding Bessel functions
 [$J_{\nu}\equiv Z^{(1)}$ for $j=1$,
 $N_{\nu}\equiv Z^{(2)}$ for $j=2$,
 $H^{(1)}_{\nu}\equiv Z^{(3)}$ for $j=3$,
 and  $H^{(2)}_{\nu}\equiv Z^{(4)}$ for $j=4$];
 specifically,
 $M_{\nu}^{(j)} (\zeta,h) \sim Z^{(j)}_{\nu} (2h \cosh \zeta)$ for the corresponding Bessel function $Z^{(j)}$
 as ${\rm Re} ( \zeta ) \sim \infty$.

 A less familiar set was introduced by Wannier~\cite{Wannier-math},
 specifically tailored for the 
 singularity and asymptotic waves
 of the seminal Vogt-Wannier paper~\cite{vogt_wannier}.
 With the suggested notational change of the comparative study in 
Refs.~\cite{IQP_Aly-Muller,IQP_Aly-Muller_2}, 
 these Mathieu-Wannier functions are denoted by
 $He^{(j)} (\zeta) \equiv he^{(j)} (i \zeta)$,
 with
 $j=1, 2$ corresponding to $u_{2,1}$
 and
 $j=3, 4$ corresponding to $u_{\pm}$ (see below).
 More precisely, the functions 
 $He^{(j)} (\zeta, q,h )$ involve a parameter $q= q(a,h)$ in lieu of $\nu$, 
 with convenient expansion 
 algorithms~\cite{park_D-brane_IQP,Muller-Kirsten_QM-book}.
 Then, reverting to the original radial variable $r$ 
 (and omitting the $q$ and $h$ parameter dependence),
 the asymptotic behaviors
for $r\sim \infty$ and $r\sim 0$ are given by
\begin{eqnarray}
\frac{1}{\sqrt{r}} \,
He^{(1)} 
(\zeta )
& \stackrel{(r \rightarrow \infty)}{\sim} &
\frac{1}{\sqrt{k}} \frac{1}{r} e^{-i\pi/4} e^{-ikr }
\label{eq:Mathieu-asymptotics_1}
\\
\frac{1}{\sqrt{r}} \,
He^{(2)} 
(\zeta )
& \stackrel{(r \rightarrow \infty)}{\sim} &
\frac{1}{\sqrt{k}} \frac{1}{r} e^{i\pi/4} e^{ikr }
\label{eq:Mathieu-asymptotics_2}
\\
\frac{1}{\sqrt{r}} \,
He^{(3)}
(\zeta )
& \stackrel{(r \rightarrow 0)}{\sim} &
\frac{1}{\lambda^{1/4}}  e^{-i\pi/4} e^{-i\sqrt{\lambda}/r }
\label{eq:Mathieu-asymptotics_3}
\\
\frac{1}{\sqrt{r}} \,
He^{(4)}
(\zeta )
& \stackrel{(r \rightarrow 0)}{\sim} &
\frac{1}{\lambda^{1/4}}  e^{i\pi/4} e^{i\sqrt{\lambda}/r }
\label{eq:Mathieu-asymptotics_4}
\; ,
\end{eqnarray}
thus uniquely specifying the bases
\begin{eqnarray}
 u_{1,2}
& = &
\sqrt{r} \, e^{\mp i \pi/2}  \, 
He^{(2,1)} (\zeta )
\label{eq:Mathieu-basis_origin}
\\
 u_{\pm}
& = &
\sqrt{r} \, e^{\pm i \pi/4} \, 
He^{(3,4)} (\zeta )
\label{eq:Mathieu-basis_infinity}
\; . 
\end{eqnarray}
As shown in Refs.~\cite{IQP_Aly-Muller,IQP_Aly-Muller_2} 
and reviewed in Refs.~\cite{park_D-brane_IQP,Muller-Kirsten_QM-book},
the conversion formulas  to the Mathieu-Bessel functions involve 
$M_{\nu}^{(3,4)} (\zeta)  \propto 
He^{(2,1)} (\zeta)
$
and 
$M_{\nu}^{(3,4)} (-\zeta)  \propto 
He^{(4,3)} (\zeta) 
$.
Then, the  transfer- and S-matrix coefficients can be found from the matching conditions 
or ``connection formulas'' for the Mathieu 
functions~\cite{Abramowitz:72,NIST,Gradshteyn-Ryzhik,Meixner-Schafke,Mathieu_McLachlan,Muller-Kirsten_QM-book}.
The 
 difficulty lies in the determination of the auxiliary parameters, which have been studied 
in some cases by the use of continued fractions and asymptotic expansions.
For example, for the Mathieu-Wannier functions~\cite{Wannier-math}
\begin{equation}
He^{(4)} =
-i e^{\Phi} 
He^{(2)} 
+
\left(
i e^{\Phi} 
\cos \pi \nu -
\cos \pi \gamma 
\right)
He^{(1)} 
\; ,
\label{eq:Wannier-Mathieu_connection}
\end{equation}
where $\Phi= \Phi (\nu,h) $ and $\gamma = \gamma (\nu,h) $ are 
related via the equations
\begin{equation}
e^{\Phi} = i \frac{ \sin \pi \gamma }{ \sin \pi \nu }
\; \; \;  
;
\; \; \;  
R \equiv e^{i\pi \gamma} \equiv \frac{  \alpha_{\nu}(h)}{ \alpha_{-\nu}(h) } = 
\frac{M^{(1)}_{-\nu} (0,h) }{ M^{(1)}_{\nu} (0,h) }
\; ,
\label{eq:Mathieu_aux-parameters}
\end{equation}
 with
$\alpha_{\nu}(h)
=
Me_{\nu}(\zeta,h^2)/M_{\nu}^{(1)} (\zeta,h)$  being the proportionality
factor for these basis functions in the connection formulas;
additional details can be gathered from the relevant 
Refs.~\cite{park_D-brane_IQP,IQP_Aly-Muller,IQP_Aly-Muller_2,Muller-Kirsten_QM-book,Wannier-math}.
Therefore, rewriting
Eq.~(\ref{eq:Wannier-Mathieu_connection})
 with the bases~(\ref{eq:Mathieu-basis_origin})--(\ref{eq:Mathieu-basis_infinity}),
one of the connection formulas reads
\begin{equation}
u_{-}
 = 
 - i e^{\Phi} 
e^{i\pi/4}
u_{1}
+
\left(
i e^{\Phi} 
\cos \pi \beta -
\cos \pi \gamma 
\right)
e^{-3i\pi/4}
u_{2}
\; ,
\label{eq:Wannier-Mathieu_connection2}
\end{equation}
which, by comparison against Eq.~(\ref{eq:basis-transformation})
(or the complex conjugate of Eq.~(\ref{eq:u_+from-u_12})),
yields the coefficients
\begin{eqnarray}
\alpha
& = &
 \left(
i e^{\Phi} 
\cos \pi \beta -
\cos \pi \gamma 
\right)
e^{-3i\pi/4}
 = 
\frac{ \sin \pi (\gamma + \nu )}{\sin \pi \nu } \, e^{i\pi/4}
\label{eq:IQP-alpha}
\\
\beta^{*}
&=&
 i e^{\Phi} e^{i\pi/4}
= 
-
\frac{ \sin \pi \gamma }{\sin \pi \nu } \, e^{i\pi/4} 
\label{eq:IQP-beta}
\; .
\end{eqnarray}
Thus,
 from Eq.~(\ref{eq:transmission-alpha_connection})
 via the reciprocal of Eq.~(\ref{eq:IQP-alpha}),
\begin{equation}
\mathcal{T}= 
\frac{\sin \pi \nu }{ \sin \pi (\gamma + \nu )}
 \, e^{- i\pi/4}
\end{equation}
and, 
from Eqs.~(\ref{eq:reflection-alpha_connection}), (\ref{eq:transmission-alpha_connection}),
and (\ref{eq:S-matrix_Mobius-phase})
 via the negative ratio of Eq.~(\ref{eq:IQP-beta}) with (\ref{eq:IQP-alpha}),
\begin{equation}
{\mathcal R'}
=
-i \,  {\mathcal R}^{*}
= 
\frac{\sin \pi \gamma }{ \sin \pi (\gamma + \nu )}
\; .
\end{equation}
Now,
Ref.~\cite{vogt_wannier} and all the other references mentioned in this subsection 
(on
absorption by the inverse quartic potential) assume 
$\Omega=0$, which is the case of total absorption; for this particular value, 
Eq.~(\ref{eq:asymptotic-S-matrix_zero-Omega})
yields
\begin{equation}
\hat{S}_{\rm asymp}(\Omega = 0)
=
{\mathcal R'}
=
 \frac{ \, \sin \pi \gamma}{\sin\pi (\gamma + \nu)}
=
 \frac{R - 1/R}{ Re^{i\pi \nu}- (Re^{i\pi \nu})^{-1}}
\; 
\end{equation}
(with $R$ defined in Eq.~(\ref{eq:Mathieu_aux-parameters})).
These formulas  agree with the known result from the literature, 
when one includes the extra phase factor 
of Eq.~(\ref{eq:S-matrix_and_reduced-S-matrix}),
with $\overline{\nu}=1/2$ for $d=3$ dimensions.
Moreover,
partial wave analysis has been carried out in great detail in several works since the original Ref.~\cite{vogt_wannier};
the absorptivity, defined by Eq.~(\ref{eq:absorptivity})
 for each angular momentum channel
 has also been studied in great detail in Refs.~\cite{park_D-brane_IQP,Muller-Kirsten_QM-book}.

\section{Conclusions}
\label{sec:conclusions}

We have derived a framework of singular interactions that includes the whole 
family of solutions for all non-contact singular potentials.
A novel ingredient of this approach is the use
of a multichannel formalism
centered on 
the S-matrix.
Our analysis has been mainly restricted to the 
 two-channel case 
(with a conveniently chosen transfer matrix),
conforming to the case of only one singular point---but extensions to multiple singular points
can be easily accommodated.

The proposed scheme singles out two distinct behaviors:
(i) the singular behavior (close to the coordinate singularity, i.e.,
$r \sim 0$): this yields singularity waves;
(ii)
the long-range tail: responsible for the coefficients of the multichannel S-matrix
(or the associated transfer matrix). 
The behavior near the singularity  is an extension of
 Wannier's original proposal of ingoing/outgoing waves,
 which we implemented for arbitrary non-contact singular potentials 
and combined with the multichannel concept.
The robustness of this approach is verified 
with the models tackled in this paper, which
include the pure conformal potential, 
extensions with various long-range tails, and the inverse quartic potential.

Interestingly, once this procedure is established, it subsumes in a unified manner
the various other approaches known to date, including renormalization
and self-adjoint extensions.
In  its two-channel form, our framework
 can accommodate a wide range of phenomenological applications---the  details of which will be discussed elsewhere.
Open problems 
that could extend the scope of this work would involve 
specific applications 
to black hole thermodynamics and the Hawking effect, D-branes, and to nanowires.

\bigskip

\noindent
{\bf Acknowledgements}

\bigskip

We are grateful to the referee for insightful comments.
This work was supported by
the National Science Foundation under Grants 
0602340 (H.E.C.) and 
0602301 (C.R.O.);
the University of San Francisco Faculty Development Fund
(H.E.C.); and ANPCyT, Argentina (L.N.E., H.F., and C.A.G.C.).

\appendix

\section{Exponential substitution and Liouville transform}
\label{sec:exponential-Liouville}

The exponential substitution
\begin{equation}
r = \frac{1}{\mu} 
\, e^{\zeta}
\end{equation}
in Eq.~(\ref{eq:singular-QM_radial})
(with $\mu$ being an inverse length parameter),
for the power-law potential 
$V({\bf r}) \stackrel{( r \rightarrow 0)}{\sim} -\lambda/r^{ p } $, 
maps the singular point from a finite position $r=0$ to $\zeta =-\infty$.
In this representation, the corresponding multichannel framework becomes a typical 
one-dimensional scattering problem relating
$\zeta =-\infty$ with $\zeta =\infty$.
To convert the equation into its normal or canonical form without first-order derivatives,
a simultaneous Liouville transformation is applied, i.e., 
the function $u$ is replaced by $w= u/\chi $,
yielding
\begin{equation}
r^{-3/2}
\left[ \ddot{w} - \frac{w}{4} \right]
+
r^{1/2}
\biggl\{
k^{2} 
-
V
-
\left[
 \left( l + \nu \right)^{2} - 1/4
      \right]/r^{2}
\biggr\}
\, 
w
=
0
\;  ,
\label{eq:singular-QM_radial-Liouville}
\end{equation}
where the
first-order derivative is eliminated for $\chi \propto \sqrt{r}$,
viz.\/,
\begin{equation}
u ( r ) = \sqrt{r} \, w(\zeta)
\end{equation}
Thus,
\begin{equation}
\ddot{w}
+
\biggl[
k^{2} \mu^{-2} \, e^{2  \zeta }
+ \lambda \mu^{ p - 2} e^{- ( p  -2) \zeta }
-
 \left( l + \nu \right)^{2} 
\biggr]
\, 
w
=
0
\;  ,
\label{eq:singular-QM_radial-exponential}
\end{equation}
where the 
dots stand for derivatives with respect to $\zeta$, i.e., $df/d\zeta = \dot{f}$.
The particular cases $ p = 2$ and $ p = 4$ 
of Section~\ref{sec:particular_cases}
can be conveniently analyzed in this representation.

Generalizations of the technique of this Appendix can be developed to accommodate a larger class of potentials
using functions of the hypergeometric type and Schwarzian 
derivatives~\cite{Derezinski:Schrodinger-eq,Liouville-Schrodinger_Milson,Forsyth,Osgood},
including Natanzon-class potentials~\cite{Natanzon}.
These techniques were used in deriving the solutions of~\ref{sec-app:S-matrix_conformally-driven}.

\section{S-matrix for conformally-driven singular systems}
\label{sec-app:S-matrix_conformally-driven}

 Two illustrative examples of singular interactions 
with long-range
behavior are the inverse squared hyperbolic sine potential
and the conformally-modified Coulomb interaction.
These examples have a interesting scale behavior involving infrared scales,
but with an ultraviolet behavior described by pure CQM---the robustness of which is explicitly verified below.
As a generic procedure,
a dimensionless equation gives solutions in terms of a dimensionless variable $\xi$
via hypergeometric and confluent hypergeometric functions;
and the relation between the bases
${\mathcal B}_{\rm sing}$
and
${\mathcal B}_{\rm asympt}$
is found from the connection formulas of these functions.

The main features of the multichannel framework for these potentials are derived and discussed below.

\subsection{Inverse-squared hyperbolic-sine potential (modified P\"{o}schl-Teller family of potentials)}
\label{sec:ISHSP}

The inverse squared sinh interaction potential 
\begin{equation}
V(r) 
=
\frac{g}{  \sinh^{2} \gamma r } 
\; 
\label{eq:ISHSP}
\end{equation}
belongs to the family of modified (hyperbolic) P\"{o}schl-Teller 
potentials~\cite{QM_Flugge,SUSY-QM_Gango,SUSY-QM_Cooper},
which typically include also an
inverse squared cosh term (see final paragraph of this Appendix subsection).
This problem
can be solved in closed form in terms of hypergeometric functions
for the nonrelativistic 
one-particle dynamics
with zero angular momentum ($l=0$) in dimensionalities $d=1$ and $d=3$;
and for any other mathematically equivalent problem.
This can be seen from 
 Eq.~(\ref{eq:singular-QM_radial})
where the singular inverse square potential 
 of Eq.~(\ref{eq:power-law_potential}) 
 (with $p=2$)
 should be replaced by the more general 
potential~(\ref{eq:ISHSP}), and the last term is zero when $l=0, \nu=1/2$
(i.e., $|\nu| = \pm 1/2$ for $d=1$ or $d=3$).

Specifically,
writing $E= k^{2} $ 
and defining
\begin{equation}
  \;  \;  \;
   \rho = \gamma r
   \; 
   ;
   \;  \;  \;
  q= \frac{k}{\gamma} 
 \; 
 ;
  \;  \;  \;  \;
  \lambda = - \frac{g}{\gamma^{2}}
   \; 
 ;
  \;  \;  \;  \;
  s = \sqrt{1/4 - \lambda}
  \; ,
  \label{eq:ISHSP-parameters}
  \end{equation}
the differential equation 
\begin{equation}
\frac{d^{2} u }{d\rho^{2} }
+
\left[
q^{2} 
-
\frac{  \left( s^{2} - 1/4 \right) 
}{
\sinh^{2} \rho  
} 
\right]
u
=
0
 \label{eq:ISHSP-equation}
\; ,
\end{equation}
after the substitutions
\begin{equation}
\xi
=
- \sinh^{2} \rho
 \; 
   ;
   \;  \;  \;
u (\rho) = \chi (\xi) \, F (\xi)
 \label{eq:ISHSP-substitution}
\; ,
\end{equation}
turns into a hypergeometric differential equation
\begin{equation}
\frac{d^{2} F }{ d \xi^{2} }
+ \frac{ \left[ c - (a_{+}+ a_{-}  + 1 ) \xi \right] }{ \xi (1-\xi) }
\,
\frac{d F }{ d \xi }
-
\frac{ a_{+}a_{-} }{\xi (1-\xi) } F
= 0
 \label{eq:ISHSP_hypergeometric-eq}
\; 
\end{equation}
when $\chi (\xi) = \left( - \xi \right)^{(s+1/2)/2}$.
Thus,  the general solution takes the form
\begin{equation}
u 
\propto
\left( - \xi \right)^{(s+1/2)/2}
\biggl( \Omega N_{+} F_{+} + N_{-}  F_{-}
 \biggr)
  \label{eq:ISHSP_general-solution}
 \; ,
\end{equation}
where 
 (from the hypergeometric connection formulas and the Legendre duplication formula 
 for the gamma function~\cite{Abramowitz:72})
\smallskip
\begin{equation}
F_{+}
= 
\begin{dcases}
{_{2}}F_{1} 
(a_{+}, a_{-}; s+1; \xi)
\;
,  
 & 
{\rm for} \;  |\xi| <1
 \\
 \mbox{\Large
$
\frac{2^{2s}  \, \Gamma (s+1) \Gamma (i q) }{\sqrt{\pi} \, \Gamma (1/2 + s + iq) } $
}
\left( 2 \sqrt{ - \xi }\right)^{-2a_{-}}
{_{2}}F_{1} 
(a_{-}, 1/2- a_{+}; - iq+1; 1/\xi)
&
 \label{eq:ISHSP_Fplus}
 \\
 + \left( iq \rightarrow -iq \right)
\; , 
&
{\rm for} \;  |\xi| >1
\end{dcases}
\end{equation}
\begin{equation}
   F_{-} 
= 
\begin{dcases}
\left( - \xi \right)^{-s}
{_{2}}F_{1} 
(1/2-a_{+}, 1/2 - a_{-}; -s+1; \xi)
 \label{eq:ISHSP_Fminus}
 \; ,
 & 
{\rm for} \;  |\xi| <1
\\
\mbox{\Large
$\frac{ \Gamma (-s+1) \Gamma (- i q) }{\sqrt{\pi} \,  \Gamma (1/2 - s - iq) } $
}
\left( 2 \sqrt{ - \xi }\right)^{-2a_{+}}
{_{2}}F_{1} 
(a_{+}, 1/2- a_{-}; iq+1; 1/\xi) 
\\
+ \left( iq \rightarrow -iq \right)
&
 {\rm for} \;  |\xi| >1
\end{dcases}
\end{equation}
with
 \begin{equation}
 a_{\pm} =
 \frac{ 1  }{2} \,  \left( 1/2 + s \pm iq \right)
 \; \; \; 
\; \; \; 
, 
\; \; \; 
\; \; \; 
 s=i\Theta
 \; 
 \label{eq:ISHSP-parameters2}
 \end{equation}
  (in the singular ``strong'' regime $\lambda \geq 1/4$). 
It should be noticed that 
 $F_{+} \equiv F_{\rm reg}$ is the ``regular'' piece 
 that survives as $\Omega \rightarrow \infty$, i.e.,  
 the one that is relevant as regular solution in the analytically continued weak-coupling regime;
 while 
 $ F_{-} \equiv F_{\rm irreg}$
  is the ``irregular'' piece that exhibits ``singular'' behavior in the weak regime.
 In addition, replacing $s$ with $-s$ effectively transforms the regular into the irregular solution.

The singularity waves are normalized in the form
\begin{equation}
u_{\pm} 
=
N_{\pm} 
\hat{u}_{\pm}
\label{eq:ISHSP_basis-normalization}
\; ,
\end{equation}
from the solutions
\begin{equation}
\hat{u}_{\pm} 
=
\left( - \xi \right)^{(1/2+s )/2} F_{\pm} (\xi) 
\stackrel{(\xi \rightarrow 0)}{\sim} 
\left( \gamma r \right)^{1/2\pm i\Theta}
 \; ,
  \label{eq:ISHSP_singularity-waves-behavior}
\end{equation}
so that, by comparison with Eq.~(\ref{eq:Wannier_waves_CQM-proper}),
\begin{equation}
N_{\pm} =
\frac{1}{\sqrt{\mu \Theta}}
\left( \frac{\mu}{\gamma} \right)^{1/2 \pm i \Theta}
 \label{eq:ISHSP_normalization-constants}
 \; .
\end{equation}
Thus, the first lines of Eqs.~(\ref{eq:ISHSP_Fplus}) and (\ref{eq:ISHSP_Fminus}),
with the additional prefactors displayed 
in Eqs.~(\ref{eq:ISHSP_basis-normalization})--(\ref{eq:ISHSP_normalization-constants}), 
represent the behavior of the singularity waves $u_{\pm}$ 
near the singular point. When analytically continued as in the second and third lines of 
Eqs.~(\ref{eq:ISHSP_Fplus})
 and (\ref{eq:ISHSP_Fminus}), 
 the two building blocks appropriate for the asymptotic waves near infinity arise; 
specifically, the second line of Eq.~(\ref{eq:ISHSP_Fplus})
displays the functional form of $u_{1}$ 
while the second line of Eq.~(\ref{eq:ISHSP_Fminus})
 displays the functional form of $u_{2}$ 
(and the third lines would generate $u_{2}$ and $u_{1}$ respectively).
The two bases can be compared by using the asymptotics of the hypergeometric function
${_{2}}F_{1} (a_{+}, a_{-}; \xi)  $, with
$\sqrt{-\xi} = \sinh \rho \sim e^{\rho}/2 $,
so that 
$\left( 2 \sqrt{-\xi} \right)^{\pm iq} \sim e^{\pm ikr}$
(with $\rho = \gamma r $ and $q= k/\gamma$).
Thus, from Eq.~(\ref{eq:u_+from-u_12})
and $u_{-}= u_{+}^{*}$, the transfer matrix coefficients are
\begin{eqnarray}
\alpha 
& = & 
\frac{ e^{-i \pi/4} }{ \sqrt{ 2 \pi  \Theta} } 
\, 
\left( \frac{ k }{ 2\mu } \right)^{ i \Theta}
\,
\frac{   
q^{ 1/2-i \Theta}
  \, \Gamma (1-i \Theta) \Gamma (-i q) 
}
{ \Gamma \left( 1/2 - i \Theta - iq \right) }
\\
\beta 
& = & 
-
\frac{ e^{-i \pi/4} }{ \sqrt{ 2 \pi  \Theta} } \, 
\,
\left( \frac{ k }{ 2\mu } \right)^{ - i \Theta}
\,
\frac{   
q^{ 1/2+i \Theta}
  \, \Gamma (1+i \Theta) \Gamma (-i q) 
}
{ \Gamma \left( 1/2 + i \Theta - iq \right) }
\; ;
\end{eqnarray}
in addition, 
${\mathcal T} = 1/\alpha$
and 
\begin{equation}
{\mathcal R} 
 = 
 \frac{\beta}{\alpha}
=
- 
\left( \frac{ k }{ 2\mu } \right)^{ - 2 i \Theta}
\,
q^{ 2i \Theta}
\,
\frac{ 
\Gamma (1 + i \Theta) 
\Gamma \left(1/2 - i \Theta - i q  \right)
}{
\Gamma (1 - i \Theta) 
 \Gamma \left(1/2 + i  \Theta - i q \right) 
 }
 \; .
\end{equation}
With these coefficients, the S-matrix $\hat{S}_{\rm asymp}$
is given by Eqs.~(\ref{eq:asymptotic-S-matrix_from-transfer-matrix})
and (\ref{eq:asymptotic-S-matrix_from-MC-S-matrix}).

The robustness of the singular CQM behavior can be verified by taking the limit $\gamma \rightarrow 0$,
which, from Eqs.~(\ref{eq:ISHSP}) and (\ref{eq:ISHSP-parameters}),
gives the conformal potential 
$V(r) 
=
-\lambda/r^2$.
This amounts to the limit $q=k/\gamma \rightarrow \infty$, which can be obtained by the asymptotic ratio of gamma functions
$\Gamma (z+a)/\Gamma (z+b) \sim z^{a-b}$ for $|z| \rightarrow \infty$  (and $|{\rm arg} (z) |<\pi$); here $z=-i q$,
 thus  (with $ {\rm arg} (-i ) =-\pi/2$), 
\begin{displaymath}
e^{-i\pi/4} \Gamma (-iq) \, q^{1/2 \mp i \Theta}/\Gamma ( 1/2 \mp i \Theta -i q)
\stackrel{(q \rightarrow \infty )}{\sim} 
  e^{\pm \pi \Theta/2}
\; ,
\end{displaymath}
reproducing the required 
coefficients~(\ref{eq:transfer-matrix_CQM1})--(\ref{eq:transfer-matrix_CQM2}).

It should be noticed that we have generalized 
the solution of the modified P\"{o}schl-Teller potential
\begin{equation}
V(r) 
=
\frac{g_{1}}{  \sinh^{2} \gamma r } 
+
\frac{g_{2}}{  \cosh^{2} \gamma r } 
\; 
\label{eq:modified_Poschl-Teller}
\end{equation}
to the strong regime of the singular $\mathrm{cosech}^{2} (\gamma r)$ piece.
 In fact, this generalized potential, including the 
$\mathrm{sech}^{2} (\gamma r)$ piece,
can be solved by the same techniques discussed above, 
 and its celebrated solution has been studied multiple times, including the recent
use of SUSY quantum mechanics techniques~\cite{SUSY-QM_Gango,SUSY-QM_Cooper}. 
In our context, the nontrivial generalization to the strong regime simply involves the same steps as above, 
with the following replacements (cf.\  the parameters~(\ref{eq:ISHSP-parameters})):
$g_{j}= - \lambda_{j} \gamma^2$, $-\lambda_{j} = s_{j}^2-1/4$ ($j=1,2$), leading to 
$1+ s \rightarrow 1 + s_{1}-s_{2} = \sigma-\tau$, where $s_{1}= \sigma -1/2$ and $s_{2} = \tau + 1/2$;
the wave function has the prefactor functions $\chi (\xi)  = ( \sinh \gamma r)^{\sigma} ( \cosh \gamma r)^{-\tau} $;
and the conformal parameter in the strong sector arises from $s_{1} = i \Theta$.

\subsection{Conformally-modified Coulomb interaction}
\label{sec:CMCI}

The interaction potential 
\begin{equation}
V(r) 
=
-\frac{\lambda}{  r^{2} } - \frac{ \gamma }{r}
\; 
\label{eq:CMCI}
\end{equation}
can be solved in closed form in terms of confluent hypergeometric functions
for the nonrelativistic 
one-particle dynamics
in any number of dimensions~\cite{dong:2004};
and for any other mathematically equivalent problem, e.g.,
it describes the near-horizon physics of spin one-half fields
in black hole backgrounds~\cite{HEC:fermion-spherical}.

The solution can be derived from Eq.~(\ref{eq:singular-QM_radial}),
where the singular inverse square potential 
 of Eq.~(\ref{eq:power-law_potential}) (with $p=2$)
 should be replaced by the more general potential~(\ref{eq:CMCI}).
Specifically,
writing $E= k^{2} $
and defining
\begin{equation}
  \;  \;  \;
  q= \frac{k}{\gamma} 
   \; 
 ;
  \;  \;  \;  \;
  s = \sqrt{(l+ \nu)^{2} - \lambda}
   \; 
 ;
  \;  \;  \;  \;
\xi
=
- 2ikr
  \; ,
  \label{eq:CMCI_parameters1}
  \end{equation}
the differential equation 
\begin{equation}
\frac{d^{2} u }{d\xi^{2} } 
+
\left[
 - \frac{1}{4}
+
 \frac{(i/2q )}{\xi} 
+
\frac{1/4 - s^2 }{\xi^2}
\right]
 u = 0
 \label{eq:CMCI_differential-eq}
\; 
\end{equation}
admits the general solution 
in terms of Whittaker functions
\begin{equation}
\hat{u}_{\pm} 
=
M_{i/2q, \pm s} (\xi)
\equiv
\xi^{1/2 \pm s} e^{-\xi/2}
\,
M 
\left(
\frac{1}{2} \pm s - \frac{i}{2q},
1 \pm 2s; \xi
\right)
\; ,
\label{eq:CMCI_solutions}
\end{equation}
where $M(a, c; \xi) \equiv {_{1}}F_{1} (a, c; \xi)$
is Kummer's confluent hypergeometric function.
The relevant parameters of the functions $M(a_{\pm}, c_{\pm}; \xi)$ above are
\begin{equation}
c_{\pm}= 1 \pm 2 s
\; \; \; 
\; \; \; 
, 
\; \; \; 
\; \; \; 
a_{\pm}= \frac{1}{2} c_{\pm} - \frac{i}{2q} = 
\frac{1}{2}  \pm s - \frac{i}{2q} 
\; .
\label{eq:CMCI_parameters2}
\end{equation}

For bookkeeping purposes, notice the replacement rule $s \rightarrow -s$ from the ``regular'' to the ``irregular'' piece.
With appropriate normalizations,
$\hat{u}_{\pm} 
=
M_{i/2q, \pm s} (\xi)$
are analytic continuations of the standard Coulomb functions 
(with $l$ replaced by a complex angular momentum).
For our purposes, we instead choose our conventional 
multichannel normalization; thus,
\begin{equation}
u 
\propto
 \Omega 
 \,
  u_{+} +  u_{-} 
=
\Omega N_{+} \hat{u}_{+} + N_{-}  \hat{u}_{-} 
 \; ,
\end{equation}
where
\begin{equation}
u_{\pm} = N_{\pm} \hat{u}_{\pm}
\; ,
\end{equation}

The generic behavior near the origin of the generalized hypergeometric functions, i.e.,
$M  \stackrel{(\xi \rightarrow 0)}{\sim} 1$ implies that
\begin{equation}
\hat{u}_{\pm}
 \stackrel{(r \rightarrow 0)}{\sim}
\left( - 2ik \right)^{\frac{1}{2} \pm i \Theta} 
 r^{\frac{1}{2} \pm i \Theta} 
\; .
\end{equation}
Thus, 
by comparison with Eq.~(\ref{eq:Wannier_waves_CQM-proper}),
the normalization factors are
\begin{equation}
N_{\pm} = \frac{ 1 }{ \sqrt{ \mu \Theta } } \,
\left( \frac{ \mu }{-2ik} \right)^{\pm i \Theta + 1/2}
\; .
\end{equation}

In addition, from the asymptotics ($r \propto |\xi| \rightarrow \infty$) of
Kummer's function, 
\begin{equation}
M(a,c; \xi)
 \stackrel{( | \xi | \rightarrow \infty)}{\sim}
 e^{-i\pi a}
 \,
 \frac{ \Gamma ( c ) }{ \Gamma ( c -a ) } 
 \,
 \xi^{-a}
 +
 \frac{ \Gamma ( c ) }{ \Gamma ( a ) } 
 \,
 e^{\xi} \, \xi^{a-c}
 \; ,
\end{equation}
one concludes that
 \begin{equation}
u_{\pm}
 \stackrel{(r \rightarrow \infty)}{\sim}
i  \sqrt{k} \, N_{\pm} e^{-\pi/4q} e^{i\pi/4} 
\,
\left[
\frac{ \Gamma (1 \pm 2i\Theta )
}{
 \Gamma (1/2 \pm i\Theta + i/2q)
}
e^{\mp \pi \Theta}
\,
{u}_{1} 
-
\frac{ \Gamma (1 \pm 2i\Theta )
}{
 \Gamma (1/2 \pm i\Theta - i/2q)
}
 {u}_{2} 
\right]
\; ,
\end{equation}
where, in agreement with the modified asymptotics of 
Eqs.~(\ref{eq:WKB_waves_infinity}) and (\ref{eq:modified_asymptotic-wave-normalization}),
\begin{equation}
 {u}_{1,2} 
 \stackrel{( r \rightarrow \infty)}{\sim} 
 \frac{1}{ \sqrt{k} }
\,
e^{\mp i \pi/4}
\,
e^{\pm i kr} 
\,
 e^{\pm i \ln (2kr)/2q }
 \; .
\end{equation}
Therefore, the transfer-matrix coefficients are
\begin{eqnarray}
\alpha 
& = & 
e^{-\pi/4q}
\,
\sqrt{\frac{1}{2 \Theta}}
\,
\left( \frac{k}{2 \mu}  \right)^{i\Theta }
\,
\frac{
2^{2i\Theta } \, \Gamma (1 - 2 i \Theta)
}
{
\Gamma (1/2 -  i \Theta- i/2q)
}
\,
e^{\pi\Theta/2}
\label{eq:CMCI_alpha}
\\
\nonumber
\\
\beta 
& = & 
-
e^{-\pi/4q}
\,
\sqrt{\frac{1}{2 \Theta}}
\,
\left( \frac{k}{2 \mu}  \right)^{-i\Theta }
\,
\frac{
2^{-2i\Theta } \, \Gamma (1 + 2 i \Theta)
}
{
\Gamma (1/2 +  i \Theta- i/2q)
}
\,
e^{-\pi\Theta/2}
\label{eq:CMCI_beta}
\; ,
\end{eqnarray}
As a result, 
\begin{eqnarray}
{\mathcal R}
& = &
-
\left( \frac{\mu}{2k}  \right)^{2 i\Theta} 
\,
\frac{ 
\Gamma (1 + 2 i \Theta ) 
\,
\Gamma (1/2 - i  \Theta -i/2q )
}{
\Gamma (1- 2 i \Theta ) 
\,
\Gamma (1/2 + i  \Theta - i/2q )
}
\,
e^{-\pi \Theta}
\label{eq:CMCI_reflection-coeff}
\\
\nonumber
\\
{\mathcal T}
& = &
e^{\pi/4q}
\,
\sqrt{2 \Theta}
\,
\left( \frac{\mu}{2 k}  \right)^{i\Theta }
\,
\frac{
  \Gamma (1/2 -  i \Theta- i/2q)
}
{
 \Gamma (1 - 2 i \Theta)
}
\,
e^{-\pi\Theta/2}
\label{eq:CMCI_transmission-coeff}
\end{eqnarray}
Here, the conformal limit $\gamma \rightarrow 0$ 
can be enforced via the Legendre duplication formula~\cite{Abramowitz:72}, 
as seen by the ratios of the gamma functions in Eqs.~(\ref{eq:CMCI_alpha})--(\ref{eq:CMCI_transmission-coeff}),
 leading again to a confirmation of Eqs.~(\ref{eq:transfer-matrix_CQM1})--(\ref{eq:transfer-matrix_CQM2}).

Several lessons
are learned from these examples.
While the inverse-squared hyperbolic-sine 
potential 
 is a short-ranged, Yukawa-like interaction asymptotically, 
 the modified Coulomb potential is a long-ranged interaction that exhibits 
 the familiar infrared problems of the ordinary $1/r$ potential.
Yet, they both illustrate the common features of the
same ultraviolet, conformally-driven physics.

\end{document}